\documentclass[11pt]{article}
\usepackage{graphicx}
\usepackage{amsmath}
\usepackage{amssymb}
\usepackage{caption2}
\begin{document}
\bibliographystyle{prsty}
\begin{center}
{\large {\bf \sc{  Analysis of the D-wave $\Sigma$-type charmed baryon states  with the QCD sum rules }}} \\[2mm]
Zhi-Gang Wang \footnote{E-mail:zgwang@aliyun.com.}, Fei Lu, Yang Liu      \\
 Department of Physics, North China Electric Power University,
Baoding 071003, P. R. China

\end{center}

\begin{abstract}
We construct the $\Sigma$-type currents to investigate   the D-wave charmed baryon states with
the QCD sum rules  systematically. The predicted masses $M=3.35^{+0.13}_{-0.18}\,\rm{GeV}$ ($3.33^{+0.13}_{-0.16}\,\rm{GeV}$), $3.34^{+0.14}_{-0.18}\,\rm{GeV}$ ($3.35^{+0.13}_{-0.16}\,\rm{GeV}$) and $3.35^{+0.12}_{-0.13}\,\rm{GeV}$ ($3.35^{+0.12}_{-0.14}\,\rm{GeV}$) for the $\Omega_c(0,2,{\frac{1}{2}}^+)$,
$\Omega_c(0,2,{\frac{3}{2}}^+)$ and $\Omega_c(0,2,{\frac{5}{2}}^+)$ states are in excellent agreement with the experimental data $ 3327.1\pm1.2 \mbox{ MeV}$ from the LHCb collaboration, and support assigning the $\Omega_c(3327)$ to be the $\Sigma$-type D-wave $\Omega_c$ state with the spin-parity $J^P={\frac{1}{2}}^+$, ${\frac{3}{2}}^+$ or ${\frac{5}{2}}^+$.
\end{abstract}



\section{Introduction}

In 2017, the LHCb collaboration investigated  the  $\Xi_c^+ K^-$ invariant  mass spectrum  with a data sample  corresponding to an integrated luminosity of  $3.3\,\rm{fb}^{-1}$, and observed five narrow excited $\Omega_c$ states,
$\Omega_c(3000)$, $\Omega_c(3050)$, $\Omega_c(3066)$, $\Omega_c(3090)$, $\Omega_c(3119)$ \cite{Omegac-Five-LHCb}, the measured masses and widths are
\begin{flalign}
 & \Omega_c(3000) : M = 3000.4 \pm 0.2 \pm 0.1 \mbox{ MeV}\, , \, \Gamma = 4.5\pm0.6\pm0.3 \mbox{ MeV} \, , \nonumber \\
 & \Omega_c(3050) : M = 3050.2 \pm 0.1 \pm 0.1 \mbox{ MeV}\, , \, \Gamma = 0.8\pm0.2\pm0.1 \mbox{ MeV} \, , \nonumber \\
 & \Omega_c(3066) : M = 3065.6 \pm 0.1 \pm 0.3 \mbox{ MeV}\, , \, \Gamma = 3.5\pm0.4\pm0.2 \mbox{ MeV} \, , \nonumber \\
 & \Omega_c(3090) : M = 3090.2 \pm 0.3 \pm 0.5 \mbox{ MeV}\, , \, \Gamma = 8.7\pm1.0\pm0.8 \mbox{ MeV} \, , \nonumber \\
 & \Omega_c(3119) : M = 3119.1 \pm 0.3 \pm 0.9 \mbox{ MeV}\, , \, \Gamma = 1.1\pm0.8\pm0.4 \mbox{ MeV} \, .
\end{flalign}
Subsequently, the Belle collaboration  confirmed  four of the five narrow states  $\Omega_c(3000)$, $\Omega_c(3050)$, $\Omega_c(3066)$ and $\Omega_c(3090)$ in the  decay mode $\Xi_c^+K^-$ using the entire Belle data sample,  which corresponds to an integrated luminosity  $980 \,\rm{fb}^{-1}$ of the $e^+e^-$ collisions \cite{Omegac-Four-Belle}.

In 2021, the LHCb collaboration observed  the $\Omega_b^- \to \Xi_c^+ K^- \pi^-$ decay for the first time using the $pp$ collision data at centre-of-mass energies of 7, 8 and 13 TeV, which corresponds  to an integrated luminosity of  $9\,\rm{fb}^{-1}$, and  confirmed the four excited $\Omega_c$ states $\Omega_c(3000)$, $\Omega_c(3050)$, $\Omega_c(3066)$ and $\Omega_c(3090)$ in the $\Xi_c^+K^-$ mass projections  with significances  larger than $5\sigma$ \cite{Omegac-Four-LHCb}.

Recently, the LHCb collaboration investigated the $\Xi_c^+ K^-$ invariant mass spectrum using proton-proton collision data corresponding to an integrated luminosity of $9\,\rm{fb}^{-1}$, and confirmed all the five previously observed excited $\Omega_c$ states, namely $\Omega_c(3000)$, $\Omega_c(3050)$, $\Omega_c(3066)$, $\Omega_c(3090)$ and $\Omega_c(3119)$ \cite{Omegac-Five-LHCb-New}. Additionally, they observed two new excited states,  $\Omega_c(3185)$ and $\Omega_c(3327)$, and acquired the masses and widths of all those states  with the highest precision up to today  \cite{Omegac-Five-LHCb-New},
\begin{flalign}
 & \Omega_c(3000) : M = 3000.44\pm0.07 \mbox{ MeV}\, , \, \Gamma = 3.83\pm0.23 \mbox{ MeV} \, , \nonumber \\
 & \Omega_c(3050) : M = 3050.18\pm0.04 \mbox{ MeV}\, , \, \Gamma = 0.67\pm0.17 \mbox{ MeV} \, , \nonumber \\
 & \Omega_c(3066) : M = 3065.63\pm0.06 \mbox{ MeV}\, , \, \Gamma = 3.79\pm0.20 \mbox{ MeV} \, , \nonumber \\
 & \Omega_c(3090) : M = 3090.16\pm0.11 \mbox{ MeV}\, , \, \Gamma = 8.48\pm0.44 \mbox{ MeV} \, , \nonumber \\
 & \Omega_c(3119) : M = 3118.98\pm0.12 \mbox{ MeV}\, , \, \Gamma = 0.60\pm0.63 \mbox{ MeV} \, , \nonumber \\
 & \Omega_c(3185) : M = 3185.1\pm1.7 \mbox{ MeV}\, , \, \Gamma = 50\pm7 \mbox{ MeV} \, , \nonumber \\
 & \Omega_c(3327) : M = 3327.1\pm1.2 \mbox{ MeV}\, , \, \Gamma = 20\pm5 \mbox{ MeV} \, .
\end{flalign}

There have been several possible assignments for those $\Omega_c$ states, such as the 1P $css$ states \cite{Chen-Omega-1P,Rosner-Omega-1P,KLWang-Omega-1P,Mathur-Omega-1P,WangZhu-Omega-1P,WZG-Omega-1P,
Cheng-Omega-1P-2S,ChenB-Omega-1P-2S,Azizi-Omega-1P-2S}, 2S $css$ states \cite{Cheng-Omega-1P-2S,ChenB-Omega-1P-2S,Azizi-Omega-1P-2S,Azizi-Omega-2S,WZG-Omega-2S}, pentaquark states \cite{Polyako-Omega-penta,PingJL-Omega-penta,WZG-Omega-penta,An-Omega-penta,WZG-WHJ-Omega-penta}, molecular states \cite{Oset-Omega-mole}. In the picture of the conventional $css$ states, the $\Omega_c(3000)$,
$\Omega_c(3050)$, $\Omega_c(3066)$, $\Omega_c(3090)$ and $\Omega_c(3119)$ lie in the region of the 1P or 2S states \cite{Chen-Omega-1P,Rosner-Omega-1P,KLWang-Omega-1P,Mathur-Omega-1P,WangZhu-Omega-1P,WZG-Omega-1P,
Cheng-Omega-1P-2S,ChenB-Omega-1P-2S,Azizi-Omega-1P-2S,Azizi-Omega-2S,WZG-Omega-2S}, while the $\Omega_c(3185)$ and $\Omega_c(3327)$ lie in the region of the 2S states \cite{YGL-3185-2S-3327-1D} and 1D states \cite{YGL-3185-2S-3327-1D,LiuX-3327-1D}, respectively.

In our previous works, we explore the 1P $\Sigma$-type charmed (bottom) baryons states, 1D $\Lambda$-type charmed baryon states, and 2S $\Lambda$-type charmed (bottom) baryon states with the QCD sum rules in details by distinguishing   the contributions of the positive parity and negative parity heavy baryon states unambiguously, and make possible assignments of the $\Omega_c(3000)$, $\Omega_c(3050)$, $\Omega_c(3066)$, $\Omega_c(3090)$, $\Omega_c(3119)$ \cite{WZG-Omega-1P},  $\Omega_b(6316)$, $\Omega_b(6330)$,
$\Omega_b(6340)$, $\Omega_b(6350)$ \cite{WZG-Omegab}, $\Lambda_c(2625)$, $\Xi_c(2815)$ \cite{Wang-2625-2815},
$\Lambda_c(2860)$, $\Lambda_c(2880)$, $\Xi_c(3055)$, $\Xi_c(3080)$ \cite{WZG-D-wave-lambda}, $\Lambda_b(6072)$, $\Lambda_b(6146)$, $\Lambda_b(6152)$, $\Xi_b(6227)$, $\Xi_b(6100)$, $\Xi_b(6327)$, $\Xi_b(6333)$ \cite{YGL-1D-CPC}, $\Lambda_b(6072)$, $\Lambda_c(2765)$, $\Xi_c(2980/2970)$ \cite{WZG-2S-CPC}.  In this work, we extend our previous works to explore  the 1D $\Sigma$-type charmed baryon  states, and try to  make possible assignments of the  $\Omega_c(3185)$ and $\Omega_c(3327)$ based on calculations via the QCD sum rules.

The article is arranged in the form:  we derive the QCD sum rules for  the D-wave charmed baryon states in Sect.2; in Sect.3, we give the  numerical results and discussions; and Sect.4 is hold for
conclusions.

\section{QCD sum rules for  the $\Sigma$-type D-wave baryon states }

Firstly, we write down  the two-point correlation functions $\Pi(p)$, $\Pi_{\alpha\beta}(p)$ and  $\Pi_{\alpha\beta\mu\nu}(p)$,
\begin{eqnarray}
\Pi(p)&=&i\int d^4x e^{ip \cdot x} \langle0|T\left\{J(x)\bar{J}(0)\right\}|0\rangle \, , \nonumber\\
\Pi_{\alpha\beta}(p)&=&i\int d^4x e^{ip \cdot x} \langle0|T\left\{J_{\alpha}(x)\bar{J}_{\beta}(0)\right\}|0\rangle \, , \nonumber\\
\Pi_{\alpha\beta\mu\nu}(p)&=&i\int d^4x e^{ip \cdot x} \langle0|T\left\{J_{\alpha\beta}(x)\bar{J}_{\mu\nu}(0)\right\}|0\rangle \, ,
\end{eqnarray}
where the interpolating currents,
\begin{eqnarray}\label{Jspin0}
J(x)&=&J^i(x)\, ,\,\,\eta^i(x)\, , \nonumber\\
J_{\alpha}(x)&=&J_\alpha^i(x)\, ,\,\,\eta_\alpha^i(x)\, , \nonumber\\
J_{\alpha\beta}(x)&=&J_{\alpha\beta}^i(x)\, ,\,\,\eta_{\alpha\beta}^i(x)\, ,
\end{eqnarray}
with $i=1$, $2$, $3$,
\begin{eqnarray}
J^1(x)&=&\varepsilon^{ijk} \left[ \partial^\mu \partial^\nu u^T_i(x) C\gamma^\rho d_j(x)+\partial^\mu  u^T_i(x) C\gamma^\rho \partial^\nu d_j(x)+\partial^\nu  u^T_i(x) C\gamma^\rho \partial^\mu d_j(x)\right.\nonumber\\
&&\left.+ u^T_i(x) C\gamma^\rho \partial^\mu \partial^\nu d_j(x)\right]\Gamma_{\mu\nu\alpha\rho}\gamma^\alpha\gamma_5 c_k(x) \, ,\nonumber \\
J^2(x)&=&\varepsilon^{ijk} \left[ \partial^\mu \partial^\nu q^T_i(x) C\gamma^\rho s_j(x)+\partial^\mu  q^T_i(x) C\gamma^\rho \partial^\nu s_j(x)+\partial^\nu  q^T_i(x) C\gamma^\rho \partial^\mu s_j(x)\right.\nonumber\\
&&\left.+ q^T_i(x) C\gamma^\rho \partial^\mu \partial^\nu s_j(x)\right]\Gamma_{\mu\nu\alpha\rho}\gamma^\alpha\gamma_5 c_k(x) \, ,\nonumber \\
J^3(x)&=&\varepsilon^{ijk} \left[ \partial^\mu \partial^\nu s^T_i(x) C\gamma^\rho s_j(x)+\partial^\mu  s^T_i(x) C\gamma^\rho \partial^\nu s_j(x)+\partial^\nu  s^T_i(x) C\gamma^\rho \partial^\mu s_j(x)\right.\nonumber\\
&&\left.+ s^T_i(x) C\gamma^\rho \partial^\mu \partial^\nu s_j(x)\right]\Gamma_{\mu\nu\alpha\rho}\gamma^\alpha\gamma_5 c_k(x) \, ,
\end{eqnarray}

\begin{eqnarray}
J_\alpha^1(x)&=&\varepsilon^{ijk} \left[ \partial^\mu \partial^\nu u^T_i(x) C\gamma^\rho d_j(x)+\partial^\mu  u^T_i(x) C\gamma^\rho \partial^\nu d_j(x)+\partial^\nu  u^T_i(x) C\gamma^\rho \partial^\mu d_j(x)\right.\nonumber\\
&&\left.+ u^T_i(x) C\gamma^\rho \partial^\mu \partial^\nu d_j(x)\right]\Gamma_{\mu\nu\alpha\rho} c_k(x) \, ,\nonumber \\
J_\alpha^2(x)&=&\varepsilon^{ijk} \left[ \partial^\mu \partial^\nu q^T_i(x) C\gamma^\rho s_j(x)+\partial^\mu  q^T_i(x) C\gamma^\rho \partial^\nu s_j(x)+\partial^\nu  q^T_i(x) C\gamma^\rho \partial^\mu s_j(x)\right.\nonumber\\
&&\left.+ q^T_i(x) C\gamma^\rho \partial^\mu \partial^\nu s_j(x)\right]\Gamma_{\mu\nu\alpha\rho} c_k(x) \, ,\nonumber \\
J_\alpha^3(x)&=&\varepsilon^{ijk} \left[ \partial^\mu \partial^\nu s^T_i(x) C\gamma^\rho s_j(x)+\partial^\mu  s^T_i(x) C\gamma^\rho \partial^\nu s_j(x)+\partial^\nu  s^T_i(x) C\gamma^\rho \partial^\mu s_j(x)\right.\nonumber\\
&&\left.+ s^T_i(x) C\gamma^\rho \partial^\mu \partial^\nu s_j(x)\right]\Gamma_{\mu\nu\alpha\rho} c_k(x) \, ,
\end{eqnarray}

\begin{eqnarray}
J_{\alpha\beta}^1(x)&=&\varepsilon^{ijk} \left[ \partial^\mu \partial^\nu u^T_i(x) C\gamma^\rho d_j(x)+\partial^\mu  u^T_i(x) C\gamma^\rho \partial^\nu d_j(x)+\partial^\nu  u^T_i(x) C\gamma^\rho \partial^\mu d_j(x)\right.\nonumber\\
&&\left.+ u^T_i(x) C\gamma^\rho \partial^\mu \partial^\nu d_j(x)\right]\Gamma_{\mu\nu\rho\alpha\beta\sigma}\gamma^\sigma\gamma_5 c_k(x) \, ,\nonumber \\
J_{\alpha\beta}^2(x)&=&\varepsilon^{ijk} \left[ \partial^\mu \partial^\nu q^T_i(x) C\gamma^\rho s_j(x)+\partial^\mu  q^T_i(x) C\gamma^\rho \partial^\nu s_j(x)+\partial^\nu  q^T_i(x) C\gamma^\rho \partial^\mu s_j(x)\right.\nonumber\\
&&\left.+ q^T_i(x) C\gamma^\rho \partial^\mu \partial^\nu s_j(x)\right]\Gamma_{\mu\nu\rho\alpha\beta\sigma}\gamma^\sigma\gamma_5 c_k(x) \, ,\nonumber \\
J_{\alpha\beta}^3(x)&=&\varepsilon^{ijk} \left[ \partial^\mu \partial^\nu s^T_i(x) C\gamma^\rho s_j(x)+\partial^\mu  s^T_i(x) C\gamma^\rho \partial^\nu s_j(x)+\partial^\nu  s^T_i(x) C\gamma^\rho \partial^\mu s_j(x)\right.\nonumber\\
&&\left.+ s^T_i(x) C\gamma^\rho \partial^\mu \partial^\nu s_j(x)\right]\Gamma_{\mu\nu\rho\alpha\beta\sigma}\gamma^\sigma\gamma_5 c_k(x) \, ,
\end{eqnarray}

\begin{eqnarray}
\eta^1(x)&=&\varepsilon^{ijk} \left[ \partial^\mu \partial^\nu u^T_i(x) C\gamma^\rho d_j(x)-\partial^\mu  u^T_i(x) C\gamma^\rho \partial^\nu d_j(x)-\partial^\nu  u^T_i(x) C\gamma^\rho \partial^\mu d_j(x)\right.\nonumber\\
&&\left.+ u^T_i(x) C\gamma^\rho \partial^\mu \partial^\nu d_j(x)\right]\Gamma_{\mu\nu\alpha\rho}\gamma^\alpha\gamma_5 c_k(x) \, ,\nonumber \\
\eta^2(x)&=&\varepsilon^{ijk} \left[ \partial^\mu \partial^\nu q^T_i(x) C\gamma^\rho s_j(x)-\partial^\mu  q^T_i(x) C\gamma^\rho \partial^\nu s_j(x)-\partial^\nu  q^T_i(x) C\gamma^\rho \partial^\mu s_j(x)\right.\nonumber\\
&&\left.+ q^T_i(x) C\gamma^\rho \partial^\mu \partial^\nu s_j(x)\right]\Gamma_{\mu\nu\alpha\rho}\gamma^\alpha\gamma_5 c_k(x) \, ,\nonumber \\
\eta^3(x)&=&\varepsilon^{ijk} \left[ \partial^\mu \partial^\nu s^T_i(x) C\gamma^\rho s_j(x)-\partial^\mu  s^T_i(x) C\gamma^\rho \partial^\nu s_j(x)-\partial^\nu  s^T_i(x) C\gamma^\rho \partial^\mu s_j(x)\right.\nonumber\\
&&\left.+ s^T_i(x) C\gamma^\rho \partial^\mu \partial^\nu s_j(x)\right]\Gamma_{\mu\nu\alpha\rho}\gamma^\alpha\gamma_5 c_k(x) \, ,
\end{eqnarray}

\begin{eqnarray}
\eta_\alpha^1(x)&=&\varepsilon^{ijk} \left[ \partial^\mu \partial^\nu u^T_i(x) C\gamma^\rho d_j(x)-\partial^\mu  u^T_i(x) C\gamma^\rho \partial^\nu d_j(x)-\partial^\nu  u^T_i(x) C\gamma^\rho \partial^\mu d_j(x)\right.\nonumber\\
&&\left.+ u^T_i(x) C\gamma^\rho \partial^\mu \partial^\nu d_j(x)\right]\Gamma_{\mu\nu\alpha\rho} c_k(x) \, ,\nonumber \\
\eta_\alpha^2(x)&=&\varepsilon^{ijk} \left[ \partial^\mu \partial^\nu q^T_i(x) C\gamma^\rho s_j(x)-\partial^\mu  q^T_i(x) C\gamma^\rho \partial^\nu s_j(x)-\partial^\nu  q^T_i(x) C\gamma^\rho \partial^\mu s_j(x)\right.\nonumber\\
&&\left.+ q^T_i(x) C\gamma^\rho \partial^\mu \partial^\nu s_j(x)\right]\Gamma_{\mu\nu\alpha\rho} c_k(x) \, ,\nonumber \\
\eta_\alpha^3(x)&=&\varepsilon^{ijk} \left[ \partial^\mu \partial^\nu s^T_i(x) C\gamma^\rho s_j(x)-\partial^\mu  s^T_i(x) C\gamma^\rho \partial^\nu s_j(x)-\partial^\nu  s^T_i(x) C\gamma^\rho \partial^\mu s_j(x)\right.\nonumber\\
&&\left.+ s^T_i(x) C\gamma^\rho \partial^\mu \partial^\nu s_j(x)\right]\Gamma_{\mu\nu\alpha\rho} c_k(x) \, ,
\end{eqnarray}

\begin{eqnarray}\label{Etaspin2}
\eta_{\alpha\beta}^1(x)&=&\varepsilon^{ijk} \left[ \partial^\mu \partial^\nu u^T_i(x) C\gamma^\rho d_j(x)-\partial^\mu  u^T_i(x) C\gamma^\rho \partial^\nu d_j(x)-\partial^\nu  u^T_i(x) C\gamma^\rho \partial^\mu d_j(x)\right.\nonumber\\
&&\left.+ u^T_i(x) C\gamma^\rho \partial^\mu \partial^\nu d_j(x)\right]\Gamma_{\mu\nu\rho\alpha\beta\sigma}\gamma^\sigma\gamma_5 c_k(x) \, ,\nonumber \\
\eta_{\alpha\beta}^2(x)&=&\varepsilon^{ijk} \left[ \partial^\mu \partial^\nu q^T_i(x) C\gamma^\rho s_j(x)-\partial^\mu  q^T_i(x) C\gamma^\rho \partial^\nu s_j(x)-\partial^\nu  q^T_i(x) C\gamma^\rho \partial^\mu s_j(x)\right.\nonumber\\
&&\left.+ q^T_i(x) C\gamma^\rho \partial^\mu \partial^\nu s_j(x)\right]\Gamma_{\mu\nu\rho\alpha\beta\sigma}\gamma^\sigma\gamma_5 c_k(x) \, ,\nonumber \\
\eta_{\alpha\beta}^3(x)&=&\varepsilon^{ijk} \left[ \partial^\mu \partial^\nu s^T_i(x) C\gamma^\rho s_j(x)-\partial^\mu  s^T_i(x) C\gamma^\rho \partial^\nu s_j(x)-\partial^\nu  s^T_i(x) C\gamma^\rho \partial^\mu s_j(x)\right.\nonumber\\
&&\left.+ s^T_i(x) C\gamma^\rho \partial^\mu \partial^\nu s_j(x)\right]\Gamma_{\mu\nu\rho\alpha\beta\sigma}\gamma^\sigma\gamma_5 c_k(x) \, ,
\end{eqnarray}
with
\begin{eqnarray}
\Gamma_{\mu\nu\alpha\beta}&=&g_{\mu\alpha}g_{\nu\beta} +g_{\mu\beta}g_{\nu\alpha}-\frac{1}{2}g_{\mu\nu}g_{\alpha\beta} \, , \nonumber\\
\Gamma_{\mu\nu\rho\alpha\beta\sigma}&=&g_{\rho\sigma}\left(g_{\mu\alpha}g_{\nu\beta} +g_{\mu\beta}g_{\nu\alpha}-\frac{1}{2}g_{\mu\nu}g_{\alpha\beta} \right)\, ,
\end{eqnarray}
$q=u,\,d$, the $i$, $j$, $k$ in the $\varepsilon^{ijk}$ are color indexes, the $C$ is the charge conjugation matrix.
We choose the currents $J(x)$, $J_\alpha(x)$ and $J_{\alpha\beta}(x)$ to interpolate the spin-parity  $J^P={\frac{1}{2}}^+$, ${\frac{3}{2}}^+$ and ${\frac{5}{2}}^+$ charmed  baryon states,  respectively.  We tentatively assign the $\Omega_c(3327)$ to be the  D-wave $\Omega_c$ state with the spin-parity $J^P={\frac{1}{2}}^+$, ${\frac{3}{2}}^+$ or ${\frac{5}{2}}^+$, the currents $J^3(x)$, $\eta^3(x)$, $J^3_\alpha(x)$, $\eta^3_\alpha(x)$,  $J^3_{\alpha\beta}(x)$ and $\eta^3_{\alpha\beta}(x)$ maybe couple potentially to the  $\Omega_c(3327)$.
In the Isospin limit, the $uuc$, $udc$ and $ddc$ baryon states have  degenerated masses, while the $usc$  and $dsc$ baryon states have  degenerated masses, we only study  the $udc$ and $usc$ baryon states for simplicity.

Now we take a short digression to explain how to construct the currents in Eqs.\eqref{Jspin0}-\eqref{Etaspin2}. We usually resort to the diquark-quark model to explore the baryon states. The attractive interaction of one-gluon exchange  favors forming diquark correlations in  color antitriplet $\overline{3}_{ c}$ \cite{One-gluon-1,One-gluon-2}.
The  diquarks $\varepsilon^{ijk} q^{T}_j C\Gamma q^{\prime}_k$ (with $q$, $q^{\prime}=u$, $d$ or $s$)
  have  five  structures, where $C\Gamma=C\gamma_5$, $C$, $C\gamma_\mu \gamma_5$,  $C\gamma_\mu $ and $C\sigma_{\mu\nu}$ for the scalar, pseudoscalar, vector, axialvector  and  tensor diquarks, respectively.  The structures
$C\gamma_\mu $ and $C\sigma_{\mu\nu}$ are symmetric (in other words, they are $\Sigma$-type states), while the structures
$C\gamma_5$, $C$ and $C\gamma_\mu \gamma_5$ are antisymmetric (in other words, they are $\Lambda$-type states).
The calculations based on the QCD sum rules  lead to the conclusion  that  the favored configurations are the $C\gamma_5$ and $C\gamma_\mu$ diquark states \cite{WangLDiquark}. For the $qq$ diquark states with $q=u$, $d$ or $s$, we have to resort to the $\Sigma$-type diquark states to satisfy the Fermi-Dirac statistic without introducing additional P-wave.
In short, we prefer the $C\gamma_\mu$ diquark states in the present work.

In the diquark-quark models, we usually denote the angular momentum between the two light quarks  by $L_\rho$, and denote the angular momentum between the  light diquark and heavy quark  by $L_\lambda$. In the case of $L_\rho=0$, we obtain  the spin-parity  $J^P=0^+$ and $1^+$ diquarks, therefore the $\Lambda$-type and $\Sigma$-type baryons, respectively \cite{Korner-PPNP}. While in the case of $(L_\rho,L_\lambda)=(2,0)$, $(0,2)$ and $(1,1)$, we can obtain copious spectrum of the D-wave charmed baryon states.
For $L_\rho=2$ and $L_{\lambda}=0$, the $qq$ diquark states have the spin-parity $J^P=3^+$,
\begin{eqnarray}
&&\varepsilon^{ijk} \left\{ \left[\partial^\mu\partial^\nu q^T_i(x) C\gamma^\rho q_j(x)-\partial^\nu q^T_i(x) C\gamma^\rho \partial^\mu q_j(x) \right]-  \left[\partial^\mu q^T_i(x) C\gamma^\rho \partial^\nu q_j(x)\right.\right.\nonumber\\
&&\left.\left.-q^T_i(x) C\gamma^\rho\partial^\mu \partial^\nu q_j(x)\right]\right\} \, ,
\end{eqnarray}
however,  the $L_\rho=1$ diquark states cannot exist due to the Fermi-Dirac statistic.
In the heavy quark limit, the $c/b$-quark is static, the $\stackrel{\leftrightarrow}{\partial}_\mu=\stackrel{\rightarrow}{\partial}_\mu-\stackrel{\leftarrow}{\partial}_\mu$ is reduced to $\stackrel{\leftarrow}{\partial}_\mu$ when it operates   on the $c/b$-quark field.
Therefore, for $L_\rho=0$ and $L_{\lambda}=2$, we acquire  the $qq$ diquark states with the spin-parity  $J^P=3^+$,
\begin{eqnarray}
\partial^\mu \partial^\nu\left[\varepsilon^{ijk}q^T_i(x) C\gamma^\rho q_j(x)\right] &=&\varepsilon^{ijk} \left[\partial^\mu\partial^\nu q^T_i(x) C\gamma^\rho q_j(x)+\partial^\mu q^T_i(x) C\gamma^\rho \partial^\nu q_j(x)\right.\nonumber\\
&&\left.+\partial^\nu q^T_i(x) C\gamma^\rho \partial^\mu q_j(x)+q^T_i(x) C\gamma^\rho\partial^\mu \partial^\nu q_j(x)\right] \, .
\end{eqnarray}
Then we classify the interpolating currents  by the quantum numbers $L_{\rho}$ and $L_{\lambda}$ ,
\begin{eqnarray}
(L_{\rho},L_{\lambda})=(0,2)\,\,&{\rm for} &  \,\,J^i(x)\, ,\,\,J^i_{\alpha}(x)\, ,\,\,J^i_{\alpha\beta}(x) \, , \nonumber\\
(L_{\rho},L_{\lambda})=(2,0)\,\,&{\rm for} & \,\,\eta^i(x)\, ,\,\,\,\,\eta^i_{\alpha}(x)\, ,\,\,\eta^i_{\alpha\beta}(x)  \, ,
\end{eqnarray}
where $i=1$, $2$, $3$.

In fact, it is difficult or impossible to construct  all currents to interpolate all the D-wave baryon states with the spin-parity $J^P={\frac{1}{2}}^+$, ${\frac{3}{2}}^+$, ${\frac{5}{2}}^+$ and ${\frac{7}{2}}^+$ in a systematic way.
In Ref.\cite{WZG-D-wave-lambda}, we explore the $\Lambda$-type D-wave baryon states with the spin-parity  $J^P={\frac{3}{2}}^+$ and ${\frac{5}{2}}^+$ in details, and explore the possible assignments of the $\Lambda_c(2860)^+$, $\Lambda_c(2880)^+$, $\Xi_c(3055)^+$, $\Xi_c(3055)^0$ and $\Xi_c(3080)^+$. Experimentally,  the $\Lambda_c(2860)^+$ and $\Lambda_c(2880)^+$ have been  observed to have the spin-parity $J^P={\frac{3}{2}}^+$ and ${\frac{5}{2}}^+$ respectively by  the  LHCb collaboration \cite{LHCb2860}. Now  we study the $\Sigma$-type D-wave charmed baryon states with the spin-parity $J^P={\frac{1}{2}}^+$, ${\frac{3}{2}}^+$ and ${\frac{5}{2}}^+$.

In general, we can choose either the partial  derivative $\partial_\mu$ or covariant derivative $D_\mu$ to construct  the interpolating currents, see Eqs.\eqref{Jspin0}-\eqref{Etaspin2}. The currents with  covariant derivative $D_\mu$ are gauge covariant/invariant, but hinders  interpreting  the $\stackrel{\leftrightarrow}{D}_\mu=\stackrel{\rightarrow}{\partial}_\mu-ig_sG_\mu-\stackrel{\leftarrow}{\partial}_\mu-ig_sG_\mu $ as  angular momentum. For example, under the gauge transformation $U_{ii^\prime}(x)$ for the quark fields   $q_i(x)$,
\begin{eqnarray}
q_i(x)&\to& U_{ii^\prime}(x)q_{i^{\prime}}(x) \, ,
\end{eqnarray}
the baryon currents without partial derivatives (or with covariant derivatives) undergo,
\begin{eqnarray}
\varepsilon^{ijk}q_i(x)q_j(x)q_k(x)&\to& U_{ii^\prime}(x)U_{jj^\prime}(x)U_{kk^\prime}(x)\,\varepsilon^{ijk}q_{i^{\prime}}(x)
q_{j^{\prime}}(x)q_{k^{\prime}}(x)\, , \nonumber\\
\varepsilon^{ijk}D_\alpha q_i(x)D_\beta q_j(x)q_k(x)&\to&U_{ii^\prime}(x)U_{jj^\prime}(x)U_{kk^\prime}(x) \,\varepsilon^{ijk}D_\alpha q_{i^{\prime}}(x)
D_\beta q_{j^{\prime}}(x)q_{k^{\prime}}(x)\, , \nonumber\\
\varepsilon^{ijk}D_\alpha D_\beta q_i(x)q_j(x)q_k(x)&\to& U_{ii^\prime}(x)U_{jj^\prime}(x)U_{kk^\prime}(x)\,\varepsilon^{ijk}D_\alpha D_\beta q_{i^{\prime}}(x)
q_{j^{\prime}}(x)q_{k^{\prime}}(x)\, ,
\end{eqnarray}
where the $i$, $j$ and $k$  are color indexes, and we have neglected  other indexes and matrixes.
 The currents with partial derivative $\partial_\mu$ are not gauge covariant, but favors  interpreting  the $\stackrel{\leftrightarrow}{\partial}_\mu=\stackrel{\rightarrow}{\partial}_\mu-\stackrel{\leftarrow}{\partial}_\mu$ as  angular momentum, furthermore, the covariant derivative $D_\mu$ leads to some hybrid components in meson or baryon states due to the gluon field $G_\mu$. For example, under the gauge transformation
 $U_{ii^\prime}(x)$,
 the baryon currents with partial derivatives undergo,
\begin{eqnarray}
\varepsilon^{ijk}\partial_\alpha q_i(x)\partial_\beta q_j(x)q_k(x)&\to& \varepsilon^{ijk}\partial_\alpha U_{ii^\prime}(x)q_{i^{\prime}}(x) \partial_\beta U_{jj^\prime}(x)q_{j^{\prime}}(x)U_{kk^\prime}(x)q_{k^{\prime}}(x)\, , \nonumber\\
&\neq&U_{ii^\prime}(x)U_{jj^\prime}(x)U_{kk^\prime}(x)\,\varepsilon^{ijk} \partial_\alpha q_{i^{\prime}}(x) \partial_\beta q_{j^{\prime}}(x)q_{k^{\prime}}(x)\, , \nonumber\\
\varepsilon^{ijk}\partial_\alpha \partial_\beta q_i(x) q_j(x)q_k(x)&\to& \varepsilon^{ijk}\partial_\alpha\partial_\beta U_{ii^\prime}(x)q_{i^{\prime}}(x)  U_{jj^\prime}(x)q_{j^{\prime}}(x)U_{kk^\prime}(x)q_{k^{\prime}}(x)\, , \nonumber\\
&\neq&U_{ii^\prime}(x)U_{jj^\prime}(x)U_{kk^\prime}(x)\,\varepsilon^{ijk} \partial_\alpha \partial_\beta q_{i^{\prime}}(x) q_{j^{\prime}}(x)q_{k^{\prime}}(x)\, .
\end{eqnarray}
In this work, we present the results with both the partial  derivatives  $\partial_\alpha$ and covariant derivatives  $D_\alpha$ for completeness.

 The currents $J(0)$, $J_\alpha(0)$ and $J_{\alpha\beta}(0)$ couple potentially to the   spin-parity $J^P={\frac{1}{2}}^+$, ${\frac{3}{2}}^+$ and  ${\frac{5}{2}}^+$  charmed baryon
 states   $B_{\frac{1}{2}}^{+}$, $B_{\frac{3}{2}}^{+}$ and   $B_{\frac{5}{2}}^{+}$, respectively. Furthermore, they also couple potentially to the  spin-parity $J^P={\frac{1}{2}}^-$, ${\frac{3}{2}}^-$ and  ${\frac{5}{2}}^-$  charmed baryon states  $B_{\frac{1}{2}}^{-}$, $B_{\frac{3}{2}}^{-}$ and   $B_{\frac{5}{2}}^{-}$, respectively, because multiplying $i\gamma_5$ changes their parity  \cite{Oka96-SB,Oka96,WangPc,WZG-Negative-P,WZG-Pcs-Old,WZG-Pcs-NPB},
\begin{eqnarray}\label{Couple-P}
\langle 0| J(0)|B_{\frac{1}{2}}^{+}(p)\rangle &=&\lambda^{+}_{\frac{1}{2}} U^{+}(p,s) \, ,  \nonumber\\
\langle 0| J_{\alpha} (0)|B_{\frac{3}{2}}^{+}(p)\rangle &=&\lambda^{+}_{\frac{3}{2}} U^{+}_\alpha(p,s) \, ,  \nonumber\\
\langle 0| J_{\alpha\beta} (0)|B_{\frac{5}{2}}^{+}(p)\rangle &=&\lambda^{+}_{\frac{5}{2}} U^{+}_{\alpha\beta}(p,s) \, ,
\end{eqnarray}
\begin{eqnarray}\label{Couple-N}
\langle 0| J (0)|B_{\frac{1}{2}}^{-}(p)\rangle &=&\lambda^{-}_{\frac{1}{2}}i\gamma_5 U^{-}(p,s) \, , \nonumber\\
\langle 0| J_{\alpha} (0)|B_{\frac{3}{2}}^{-}(p)\rangle &=&\lambda^{-}_{\frac{3}{2}}i\gamma_5 U^{-}_{\alpha}(p,s) \, , \nonumber\\
\langle 0| J_{\alpha\beta} (0)|B_{\frac{5}{2}}^{-}(p)\rangle &=&\lambda^{-}_{\frac{5}{2}}i\gamma_5 U^{-}_{\alpha\beta}(p,s) \, ,
\end{eqnarray}
where the $\lambda^{\pm}_{\frac{1}{2}}$, $\lambda^{\pm}_{\frac{3}{2}}$ and $\lambda^{\pm}_{\frac{5}{2}}$ are the pole residues, the $U^{\pm}(p,s)$, $U^{\pm}_\alpha(p,s)$ and $U^{\pm}_{\alpha\beta}(p,s)$ are the Dirac and  Rarita-Schwinger spinors.

Then  we  insert  a complete set  of intermediate charmed baryon  states with the
same quantum numbers as the currents  $J(x)$, $i\gamma_5 J(x)$,  $J_\alpha(x)$,
$i\gamma_5 J_\alpha(x)$, $J_{\alpha\beta}(x)$ and
$i\gamma_5 J_{\alpha\beta}(x)$ into the correlation functions $\Pi(p)$,
$\Pi_{\alpha\beta}(p)$ and $\Pi_{\alpha\beta\mu\nu}(p)$ (according to  Eqs.\eqref{Couple-P}-\eqref{Couple-N}) to obtain the hadronic representation
\cite{SVZ79,PRT85}. We isolate the pole terms of the lowest
 charmed  baryon states with  positive/negative parity,    and reach the  results:
\begin{eqnarray}
\Pi(p) & = & {\lambda^{+}_{\frac{1}{2}}}^2  {\!\not\!{p}+ M_{+} \over M_{+}^{2}-p^{2}  } +  {\lambda^{-}_{\frac{1}{2}}}^2  {\!\not\!{p}- M_{-} \over M_{-}^{2}-p^{2}  } +\cdots  \, ,\nonumber\\
&=&\Pi_{\frac{1}{2}}(p^2)\, ,
\end{eqnarray}
\begin{eqnarray}
\Pi_{\alpha\beta}(p) & = & {\lambda^{+}_{\frac{3}{2}}}^2  {\!\not\!{p}+ M_{+} \over M_{+}^{2}-p^{2}  } \left(- g_{\alpha\beta}+\frac{\gamma_\alpha\gamma_\beta}{3}+\frac{2p_\alpha p_\beta}{3p^2}-\frac{p_\alpha\gamma_\beta-p_\beta \gamma_\alpha}{3\sqrt{p^2}}
\right)\nonumber\\
&&+  {\lambda^{-}_{\frac{3}{2}}}^2  {\!\not\!{p}- M_{-} \over M_{-}^{2}-p^{2}  } \left(- g_{\alpha\beta}+\frac{\gamma_\alpha\gamma_\beta}{3}+\frac{2p_\alpha p_\beta}{3p^2}-\frac{p_\alpha\gamma_\beta-p_\beta \gamma_\alpha}{3\sqrt{p^2}}
\right) +\cdots  \, ,\nonumber\\
&=&\Pi_{\frac{3}{2}}(p^2)\,\left(- g_{\alpha\beta}\right)+\cdots\, ,
\end{eqnarray}
\begin{eqnarray}
\Pi_{\alpha\beta\mu\nu}(p) & = & {\lambda^{+}_{\frac{5}{2}}}^2  {\!\not\!{p}+ M_{+} \over M_{+}^{2}-p^{2}  } \left[\frac{ \widetilde{g}_{\mu\alpha}\widetilde{g}_{\nu\beta}+\widetilde{g}_{\mu\beta}\widetilde{g}_{\nu\alpha}}{2}-\frac{\widetilde{g}_{\mu\nu}\widetilde{g}_{\alpha\beta}}{5}
-\frac{1}{10}\left( \gamma_{\alpha}\gamma_{\mu}+\cdots\right)\widetilde{g}_{\nu\beta}
+\cdots\right]\nonumber\\
&&+   {\lambda^{-}_{\frac{5}{2}}}^2  {\!\not\!{p}- M_{-} \over M_{-}^{2}-p^{2}  }  \left[\frac{ \widetilde{g}_{\mu\alpha}\widetilde{g}_{\nu\beta}+\widetilde{g}_{\mu\beta}\widetilde{g}_{\nu\alpha}}{2}-\frac{\widetilde{g}_{\mu\nu}\widetilde{g}_{\alpha\beta}}{5}
+\cdots\right]   +\cdots \, , \nonumber\\
&=&\Pi_{\frac{5}{2}}(p^2)\,\frac{ g_{\mu\alpha}g_{\nu\beta}+g_{\mu\beta}g_{\nu\alpha}}{2}+\cdots \, ,
\end{eqnarray}
where $\widetilde{g}_{\mu\nu}=g_{\mu\nu}-\frac{p_{\mu}p_{\nu}}{p^2}$,  and we have used
 summations over the polarizations $s$ in the spinors $U^{\pm}(p,s)$, $U^{\pm}_\alpha(p,s)$ and $U^{\pm}_{\alpha\beta}(p,s)$  \cite{HuangShiZhong},
\begin{eqnarray}
\sum_s U \overline{U}&=&\!\not\!{p}+M_{\pm} \,  , \nonumber\\
\sum_s U_\alpha \overline{U}_\beta&=&\left(\!\not\!{p}+M_{\pm}\right)\left( -g_{\alpha\beta}+\frac{\gamma_\alpha\gamma_\beta}{3}+\frac{2p_\alpha p_\beta}{3p^2}-\frac{p_\alpha\gamma_\beta-p_\beta \gamma_\alpha}{3\sqrt{p^2}} \right) \,  , \nonumber \\
\sum_s U_{\mu\nu}\overline {U}_{\alpha\beta}&=&\left(\!\not\!{p}+M_{\pm}\right)\left\{\frac{\widetilde{g}_{\mu\alpha}\widetilde{g}_{\nu\beta}+\widetilde{g}_{\mu\beta}\widetilde{g}_{\nu\alpha}}{2} -\frac{\widetilde{g}_{\mu\nu}\widetilde{g}_{\alpha\beta}}{5}-\frac{1}{10}\left( \gamma_{\mu}\gamma_{\alpha}+\frac{\gamma_{\mu}p_{\alpha}-\gamma_{\alpha}p_{\mu}}{\sqrt{p^2}}-\frac{p_{\mu}p_{\alpha}}{p^2}\right)\widetilde{g}_{\nu\beta}\right. \nonumber\\
&&-\frac{1}{10}\left( \gamma_{\nu}\gamma_{\alpha}+\frac{\gamma_{\nu}p_{\alpha}-\gamma_{\alpha}p_{\nu}}{\sqrt{p^2}}-\frac{p_{\nu}p_{\alpha}}{p^2}\right)\widetilde{g}_{\mu\beta}
-\frac{1}{10}\left( \gamma_{\mu}\gamma_{\beta}+\frac{\gamma_{\mu}p_{\beta}-\gamma_{\beta}p_{\mu}}{\sqrt{p^2}}-\frac{p_{\mu}p_{\beta}}{p^2}\right)\widetilde{g}_{\nu\alpha}\nonumber\\
&&\left.-\frac{1}{10}\left( \gamma_{\nu}\gamma_{\beta}+\frac{\gamma_{\nu}p_{\beta}-\gamma_{\beta}p_{\nu}}{\sqrt{p^2}}-\frac{p_{\nu}p_{\beta}}{p^2}\right)\widetilde{g}_{\mu\alpha} \right\} \, ,
\end{eqnarray}
and $p^2=M^2_{\pm}$ on  mass-shell.

On the other hand, the currents $J_\alpha(0)$ and $J_{\alpha\beta}(0)$ have substantial  couplings with  the spin-parity $J^P={\frac{1}{2}}^\pm$ and $J^P={\frac{1}{2}}^\pm$, ${\frac{3}{2}}^\pm$ charmed baryon states, respectively,  we choose the tensor structures $g_{\alpha\beta}$ and $g_{\mu\alpha}g_{\nu\beta}+g_{\mu\beta}g_{\nu\alpha}$ to explore the spin-parity $J^P={\frac{3}{2}}^+$ and ${\frac{5}{2}}^+$ baryon states, respectively, the corresponding $J^P={\frac{1}{2}}^\pm$, ${\frac{3}{2}}^\pm$ baryon states cannot  contaminate the QCD sum rules \cite{WangPc}.

Now we obtain the hadronic spectral densities  through dispersion relation,
\begin{eqnarray}
\frac{{\rm Im}\Pi_{j}(s)}{\pi}&=&\!\not\!{p} \left[{\lambda^{+}_{j}}^2 \delta\left(s-M_{+}^2\right)+{\lambda^{-}_{j}}^2 \delta\left(s-M_{-}^2\right)\right] +\left[M_{+}{\lambda^{+}_{j}}^2 \delta\left(s-M_{+}^2\right)-M_{-}{\lambda^{-}_{j}}^2 \delta\left(s-M_{-}^2\right)\right]\, , \nonumber\\
&=&\!\not\!{p}\, \rho^1_{j,H}(s)+\rho^0_{j,H}(s) \, ,
\end{eqnarray}
where $j=\frac{1}{2}$, $\frac{3}{2}$, $\frac{5}{2}$, we add the subscript $H$ to stand for  the hadron side,
then we introduce the weight  function $\exp\left(-\frac{s}{T^2}\right)$ to suppress the higher  resonances (excited states) and continuum states  to achieve the QCD sum rules at the hadron side,
\begin{eqnarray}
\int_{m_c^2}^{s_0}ds \left[\sqrt{s}\rho^1_{j,H}(s)+\rho^0_{j,H}(s)\right]\exp\left( -\frac{s}{T^2}\right)
&=&2M_{+}{\lambda^{+}_{j}}^2\exp\left( -\frac{M_{+}^2}{T^2}\right) \, ,
\end{eqnarray}
where the $s_0$ are the continuum thresholds and the $T^2$ are the Borel parameters \cite{WZG-Omega-1P,WZG-Omegab,Wang-2625-2815,WZG-D-wave-lambda,YGL-1D-CPC,WZG-2S-CPC,WangPc,WZG-Negative-P,WZG-Pcs-Old}. Because of
the special combination $\sqrt{s}\rho^1_{j,H}(s)+\rho^0_{j,H}(s)$, the negative parity charmed baryon states cannot contaminate the QCD sum rules, and they saturate other QCD sum rules unambiguously,
 \begin{eqnarray}
\int_{m_c^2}^{s_0}ds \left[\sqrt{s}\rho^1_{j,H}(s)-\rho^0_{j,H}(s)\right]\exp\left( -\frac{s}{T^2}\right)
&=&2M_{-}{\lambda^{-}_{j}}^2\exp\left( -\frac{M_{-}^2}{T^2}\right) \, .
\end{eqnarray}

At the QCD side, we  calculate the correlation functions $\Pi(p)$,
 $\Pi_{\alpha\beta}(p)$ and $\Pi_{\alpha\beta\mu\nu}(p)$ with the full light quark propagators $S_{ij}(x)$,
 \begin{eqnarray}\label{Sij-x}
S_{ij}(x)&=& \frac{i\delta_{ij}\!\not\!{x}}{ 2\pi^2x^4}
-\frac{\delta_{ij}m_q}{4\pi^2x^2}-\frac{\delta_{ij}\langle
\bar{q}q\rangle}{12} +\frac{i\delta_{ij}\!\not\!{x}m_q
\langle\bar{q}q\rangle}{48}-\frac{\delta_{ij}x^2\langle \bar{q}g_s\sigma Gq\rangle}{192}\nonumber\\
&&+\frac{i\delta_{ij}x^2\!\not\!{x} m_q\langle \bar{q}g_s\sigma
 Gq\rangle }{1152} -\frac{ig_s G^{a}_{\alpha\beta}t^a_{ij}(\!\not\!{x}
\sigma^{\alpha\beta}+\sigma^{\alpha\beta} \!\not\!{x})}{32\pi^2x^2} -\frac{1}{8}\langle\bar{q}_j\sigma^{\mu\nu}q_i \rangle \sigma_{\mu\nu} +\cdots \, , \nonumber \\
\end{eqnarray}
 and  full $c$-quark propagator $C_{ij}(x)$,
\begin{eqnarray}
C_{ij}(x)&=&\frac{i}{(2\pi)^4}\int d^4k e^{-ik \cdot x} \left\{
\frac{\delta_{ij}}{\!\not\!{k}-m_c}
-\frac{g_sG^n_{\alpha\beta}t^n_{ij}}{4}\frac{\sigma^{\alpha\beta}(\!\not\!{k}+m_c)+(\!\not\!{k}+m_c)
\sigma^{\alpha\beta}}{(k^2-m_c^2)^2}\right.\nonumber\\
&&\left. -\frac{g_s^2 (t^at^b)_{ij} G^a_{\alpha\beta}G^b_{\mu\nu}(f^{\alpha\beta\mu\nu}+f^{\alpha\mu\beta\nu}+f^{\alpha\mu\nu\beta}) }{4(k^2-m_c^2)^5}+\cdots\right\} \, , \end{eqnarray}
\begin{eqnarray}
f^{\alpha\beta\mu\nu}&=&(\!\not\!{k}+m_c)\gamma^\alpha(\!\not\!{k}+m_c)\gamma^\beta(\!\not\!{k}+m_c)\gamma^\mu(\!\not\!{k}+m_c)\gamma^\nu(\!\not\!{k}+m_c)\, ,
\end{eqnarray}
 $q=u,d,s$,  $t^n=\frac{\lambda^n}{2}$, the $\lambda^n$ is the Gell-Mann matrix \cite{PRT85}. In Eq.\eqref{Sij-x}, we adopt the $\langle\bar{q}_j\sigma_{\mu\nu}q_i \rangle$  comes from the Fierz transformation of the $\langle q_i \bar{q}_j\rangle$ to  absorb the gluons  emitted from the other  quark lines to  extract the mixed condensate  $\langle\bar{q}g_s\sigma G q\rangle$.
Then we accomplish  all the integrals in the coordinate and momentum spaces  in sequence to achieve the QCD representation up to the vacuum condensates of dimension 10 in a consistent way \cite{WZG-Omega-1P,WZG-Omegab,Wang-2625-2815,WZG-D-wave-lambda,YGL-1D-CPC,WZG-2S-CPC,WZG-Negative-P}, and  achieve   the QCD spectral densities  through  dispersion relation,
\begin{eqnarray}
\frac{{\rm Im}\Pi_{j}(s)}{\pi}&=&\!\not\!{p}\, \rho^1_{j,QCD}(s)+\rho^0_{j,QCD}(s) \, ,
\end{eqnarray}
where $j=\frac{1}{2}$, $\frac{3}{2}$, $\frac{5}{2}$. For simplicity, we give the explicit expressions of the QCD spectral densities in the Appendix.

\begin{figure}
 \centering
  \includegraphics[totalheight=3cm,width=10cm]{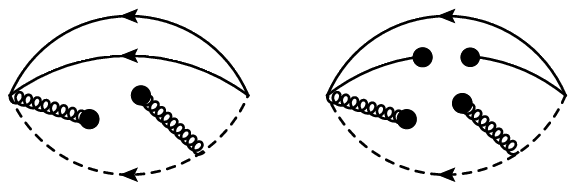}\\
  \vspace{1cm}
   \includegraphics[totalheight=3cm,width=10cm]{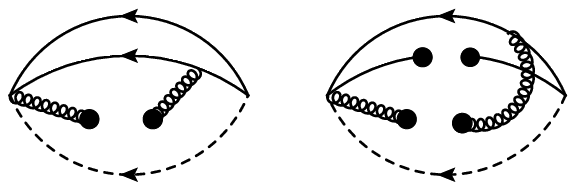}\\
  \vspace{1cm}
   \includegraphics[totalheight=3cm,width=10cm]{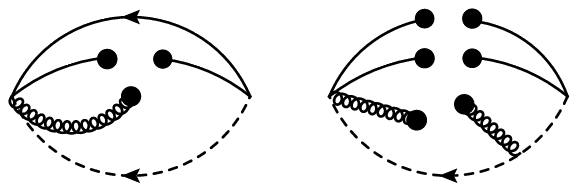}\\
   \vspace{1cm}
  \includegraphics[totalheight=3cm,width=10cm]{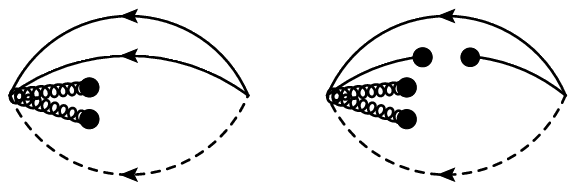}\\
  \vspace{1cm}
  \includegraphics[totalheight=3cm,width=5cm]{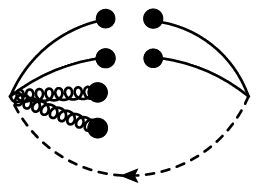}\\
       \caption{ The additional Feynman diagrams (originate from the covariant derivatives) make contributions to the vacuum condensates, where the dashed (solid) lines denote the heavy (light) quark lines.}\label{Feyn-Dr-gg}
\end{figure}

\begin{figure}
 \centering
    \includegraphics[totalheight=4cm,width=12cm]{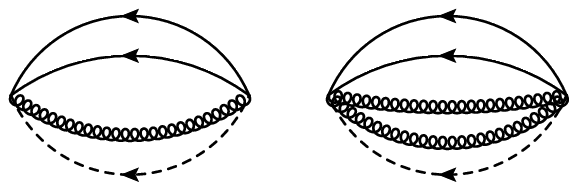}
   \caption{ The additional Feynman diagrams (originate from the covariant derivatives) make contributions of the orders $\mathcal{O}(\alpha_s)$ and $\mathcal{O}(\alpha_s^2)$, where the dashed (solid) lines denote the heavy (light) quark lines.}\label{Feyn-Dr-per}
\end{figure}

In calculations, we carry out the operator product expansion by choosing the partial derivatives $\partial_\mu$ firstly, then take the simple replacement $\delta_{ij}\partial_\mu \to \delta_{ij}\partial_\mu-ig_sG^a_\mu \frac{\lambda^a_{ij}}{2}$ in the vertexes to take account of the additional terms originate from the covariant derivatives. In Fig.\ref{Feyn-Dr-gg}, we show the additional Feynman diagrams (originate from the covariant derivatives) make contributions to the vacuum condensates $\langle \bar{q}g_s\sigma Gq\rangle $,
$\langle \frac{\alpha_sGG}{\pi}\rangle$, $\langle \bar{q}q\rangle \langle \frac{\alpha_sGG}{\pi}\rangle$, $\langle \bar{q}g_s\sigma Gq\rangle^2 $ and $\langle \bar{q}q\rangle^2 \langle \frac{\alpha_sGG}{\pi}\rangle$ with $q=u$, $d$ or $s$, which correspond to  truncations of the quark-gluon operators of the orders $\mathcal{O}(\alpha_s^k)$ with $k\leq 1$ (adopted in all our previous works) \cite{WZG-Omega-1P,WZG-Omegab,Wang-2625-2815,WZG-D-wave-lambda,YGL-1D-CPC,WZG-2S-CPC,WZG-Negative-P}. We have chosen the fixed-point  gauge $G_\mu(x)=\frac{1}{2}x^\sigma G_{\sigma\mu}(0)$, and the vacuum condensates $\langle \bar{q}g_s\sigma Gq\rangle^2 $ and $\langle \bar{q}q\rangle^2 \langle \frac{\alpha_sGG}{\pi}\rangle$ happen to have no contribution. In Fig.\ref{Feyn-Dr-per}, we show the
perturbative contributions originate from the gluons in the covariant derivatives, they are of the orders $\mathcal{O}(\alpha_s)$ and $\mathcal{O}(\alpha_s^2)$ and are neglected, just like what have been done  in the literatures \cite{ChenHX-Omega-D-wave}. Furthermore, taking account of the diagrams in Fig.\ref{Feyn-Dr-per} amounts to introducing some valence gluon Fock components in the interpolating currents, while we choose the covariant derivative only for the sake of obtaining gauge covariant/invariant currents.

Now  we  suppose quark-hadron duality below the continuum thresholds  $s_0$, again we resort to the weight  function $\exp\left(-\frac{s}{T^2}\right)$ to suppress the higher resonances (excited states) and continuum states to achieve   the  QCD sum rules:
\begin{eqnarray}\label{QCDSR}
2M_{+}{\lambda^{+}_{j}}^2\exp\left( -\frac{M_{+}^2}{T^2}\right)
&=& \int_{m_c^2}^{s_0}ds \left[\sqrt{s}\rho^1_{j,QCD}(s)+\rho^0_{j,QCD}(s)\right]\exp\left( -\frac{s}{T^2}\right)\, .
\end{eqnarray}

We differentiate  Eq.\eqref{QCDSR} in regard  to  $\frac{1}{T^2}$, then eliminate the
 pole residues $\lambda^{+}_j$ through a fraction, and achieve  the QCD sum rules for
 the $\Sigma$-type D-wave  charmed baryon  masses,
 \begin{eqnarray}\label{SR-mass}
 M^2_{+} &=& \frac{-\frac{d}{d(1/T^2)}\int_{m_c^2}^{s_0}ds \left[\sqrt{s}\rho^1_{j,QCD}(s)+\rho^0_{j,QCD}(s)\right]\exp\left( -\frac{s}{T^2}\right)}{\int_{m_c^2}^{s_0}ds \left[\sqrt{s}\rho^1_{j,QCD}(s)+\rho^0_{j,QCD}(s)\right]\exp\left( -\frac{s}{T^2}\right)}\, .
\end{eqnarray}

For the Ioffe currents, we can obtain the relation,
\begin{eqnarray}
M_N&\approx&-\frac{8\pi^2\langle \bar{q}q\rangle}{T^2}\, ,
\end{eqnarray}
by neglecting the tiny $u$ and $d$ quark masses for the proton and neutron \cite{PRT85}, which indicates that the
ground state masses mainly originate from the quark condensates. Such simple relation does not exist in the present case due to the large  $c$-quark mass and additional D-wave, the net effects of the perturbative terms, quark condensates, gluon condensates and mixed condensates lead to the excited baryon masses.

\section{Numerical results and discussions}
At the QCD side, we take the standard vacuum condensates
$\langle\bar{q}q \rangle=-(0.24\pm 0.01\, \rm{GeV})^3$,  $\langle\bar{s}s \rangle=(0.8\pm0.1)\langle\bar{q}q \rangle$,
 $\langle\bar{q}g_s\sigma G q \rangle=m_0^2\langle \bar{q}q \rangle$, $\langle\bar{s}g_s\sigma G s \rangle=m_0^2\langle \bar{s}s \rangle$,
$m_0^2=(0.8 \pm 0.1)\,\rm{GeV}^2$, $\langle \frac{\alpha_s
GG}{\pi}\rangle=0.012\pm0.004\,\rm{GeV}^4$    at the energy scale  $\mu=1\, \rm{GeV}$
\cite{SVZ79,PRT85,ColangeloReview}, and  take the $\overline{MS}$ masses $m_{c}(m_c)=(1.275\pm0.025)\,\rm{GeV}$ and $m_s(\mu=2\,\rm{GeV})=(0.095\pm0.005)\,\rm{GeV}$
 from the Particle Data Group \cite{PDG}.
In addition,  we take  account of
the energy-scale dependence of  all the input  parameters, such as the vacuum condensates and $\overline{MS}$ masses,
 \begin{eqnarray}
 \langle\bar{q}q \rangle(\mu)&=&\langle\bar{q}q\rangle({\rm 1 GeV})\left[\frac{\alpha_{s}({\rm 1 GeV})}{\alpha_{s}(\mu)}\right]^{\frac{12}{33-2n_f}}\, , \nonumber\\
 \langle\bar{s}s \rangle(\mu)&=&\langle\bar{s}s \rangle({\rm 1 GeV})\left[\frac{\alpha_{s}({\rm 1 GeV})}{\alpha_{s}(\mu)}\right]^{\frac{12}{33-2n_f}}\, , \nonumber\\
 \langle\bar{q}g_s \sigma Gq \rangle(\mu)&=&\langle\bar{q}g_s \sigma Gq \rangle({\rm 1 GeV})\left[\frac{\alpha_{s}({\rm 1 GeV})}{\alpha_{s}(\mu)}\right]^{\frac{2}{33-2n_f}}\, ,\nonumber\\
  \langle\bar{s}g_s \sigma Gs \rangle(\mu)&=&\langle\bar{s}g_s \sigma Gs \rangle({\rm 1 GeV})\left[\frac{\alpha_{s}({\rm 1 GeV})}{\alpha_{s}(\mu)}\right]^{\frac{2}{33-2n_f}}\, ,\nonumber\\
m_c(\mu)&=&m_c(m_c)\left[\frac{\alpha_{s}(\mu)}{\alpha_{s}(m_c)}\right]^{\frac{12}{33-2n_f}} \, ,\nonumber\\
m_s(\mu)&=&m_s({\rm 2GeV} )\left[\frac{\alpha_{s}(\mu)}{\alpha_{s}({\rm 2GeV})}\right]^{\frac{12}{33-2n_f}}\, ,\nonumber\\
\alpha_s(\mu)&=&\frac{1}{b_0t}\left[1-\frac{b_1}{b_0^2}\frac{\log t}{t} +\frac{b_1^2(\log^2{t}-\log{t}-1)+b_0b_2}{b_0^4t^2}\right]\, ,
\end{eqnarray}
  where $t=\log \frac{\mu^2}{\Lambda^2}$, $b_0=\frac{33-2n_f}{12\pi}$, $b_1=\frac{153-19n_f}{24\pi^2}$, $b_2=\frac{2857-\frac{5033}{9}n_f+\frac{325}{27}n_f^2}{128\pi^3}$,  $\Lambda=213\,\rm{MeV}$, $296\,\rm{MeV}$  and  $339\,\rm{MeV}$ for the flavors  $n_f=5$, $4$ and $3$, respectively  \cite{PDG,Narison-mix}.
We explore the $\Sigma$-type D-wave charmed baryon states and choose the flavor numbers $n_f=4$.

\begin{figure}
 \centering
 \includegraphics[totalheight=7cm,width=10cm]{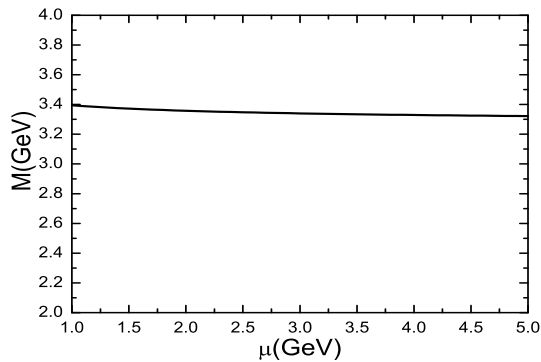}
        \caption{ The mass  of the $\Omega_c(0,2,{\frac{3}{2}}^+)$ state with variations of the energy scale $\mu$ for the central values of the iuput parameters (for the current with covariant derivatives).  }\label{Omega-GG-mu}
\end{figure}

As the energy scales of the QCD spectral densities are concerned, we give some discussions.  In the heavy quark limit, the $Q$-quark plays a role as a static well potential in the $qq^\prime Q$, $q\bar{q}^{\prime }Q\overline{Q}$, $qq^{\prime }q^{\prime \prime }Q \overline{Q}$ and $\bar{q}\bar{q}^{\prime}QQ$ systems,
then we introduce the effective heavy quark masses $\mathbb{M}_Q$ and divide the baryon/multiquark states into both the heavy and light degrees of freedom. If we neglect the tiny $u$ and $d$ quark masses, we acquire the heavy degrees of freedom
${\mathbb{M}}_Q/2{\mathbb{M}}_Q$ and light degrees of freedom $V=\sqrt{M^2_{B/X/Y/Z/T/P}-({\mathbb{M}}_Q/2{\mathbb{M}}_Q)^2}$ (or virtuality), then we set the $V$ to be the energy scales $\mu$ of the QCD spectral densities, therefore achieve  the energy scale formula, where the $B$, $X$, $Y$, $Z$, $T$ and $P$ stand for the traditional baryon states, tetraquark (molecular) states and pentaquark states, respectively   \cite{WangPc,WangTetraquark}. In addition, we take  account of  the light flavor $SU_f(3)$ breaking effects by introducing  the effective $s$-quark mass $\mathbb{M}_s$ to achieve the modified energy scale formula $\mu=\sqrt{M^2_{B/X/Y/Z/T/P}-({\mathbb{M}}_Q/2{\mathbb{M}}_Q)^2}-k\mathbb{M}_s$, where $k=0$, $1$, $2$, $3$ count for the numbers of the $s$-quark, which works well \cite{WangIJMPA-mole,XinQ-EPJA}.
The effective quark masses ${\mathbb{M}}_{Q/s}$ have  universal values,
 we take the updated effective $c$-quark mass ${\mathbb{M}}_c=1.82\,\rm{GeV}$ \cite{WangEPJC4260} and effective $s$-quark mass ${\mathbb{M}}_s=0.20\,\rm{GeV}$ \cite{WangIJMPA-mole,XinQ-EPJA}, then if we identify  the $\Omega_c(3327)$ as the traditional D-wave $\Sigma$-type baryon state, we achieve  the suitable  energy scale $\mu=\sqrt{M^2_{\Omega}-{\mathbb{M}}_c^2}-2\mathbb{M}_s=2.4\,\rm{GeV}$ for the QCD spectral densities. In calculations, we observe that the modified energy scale formula leads to the universal energy scales  $\mu=2.4\,\rm{GeV}$ ($2.7\,\rm{GeV}$) approximately for the $(L_\rho,L_\lambda)=(0,2)$ ($(2,0)$) $\Sigma$-type charmed baryon states. In fact, at the energy scales $\mu\geq 2.0\,\rm{GeV}$, the predicted baryon masses change very slowly with variations of the energy scales of the QCD spectral densities, it is reasonable to choose the energy scales $\mu=2.4\,\rm{GeV}$ and $2.7\,\rm{GeV}$. For example, in Fig.\ref{Omega-GG-mu}, we plot the mass  of the $\Omega_c(0,2,{\frac{3}{2}}^+)$ state with variations of the energy scale $\mu$ for the central values of the iuput parameters (for the current with covariant derivatives, see Table \ref{Borel-GG}). From the figure, we can see explicitly that the predicted mass decreases slowly with increase of the energy scale of the QCD spectral density, slightly larger or smaller energy scales cannot change the predictions remarkably.

  We search for the ideal Borel parameters $T^2$ and continuum threshold parameters $s_0$  to satisfy the two basic criteria to achieve reliable  QCD sum rules: firstly, pole dominance at the hadron side, we place universal restrictions on  the pole contributions, about $(40-85)\%$; secondly, convergence of the operator product expansion at the QCD side, as the dominant contributions come from the perturbative terms,  such a criterion is easy to satisfy. In the present work, we calculate the vacuum condensates up to dimension 10 in a consistent way, the higher dimensional vacuum condensates play an important role in acquiring    the Borel windows, while in the Borel windows, they play a minor role, for example, the contributions of the vacuum condensates of dimension 10 are about $(1-3)\%$, $(1-4)\%$ and $(1-2)\%$ for the $\Omega_c(0,2,{\frac{1}{2}}^+)$, $\Omega_c(0,2,{\frac{3}{2}}^+)$ and $\Omega_c(0,2,{\frac{5}{2}}^+)$ states, respectively,  for the currents with the covariant derivatives.

 Finally, we acquire  suitable Borel parameters $T^2$, continuum threshold parameters $s_0$,  pole contributions and perturbative contributions, see Tables \ref{Borel}-\ref{Borel-GG}, where we choose uniform Borel windows, $T^2_{max}-T^2_{min}=0.6\,\rm{GeV}^2$, the subscripts "max" and "min" stand for the maximum and minimum values respectively.  We take  account of all uncertainties  of the input   parameters, and obtain the values of the masses and pole residues of the ground state D-wave charmed  baryon states, which are shown in  Tables \ref{mass}-\ref{mass-GG}. From Tables \ref{Borel}, \ref{mass} and Tables \ref{Borel-GG}, \ref{mass-GG}, we can see clearly that the continuum threshold parameters and predicted baryon masses  have the relation $\sqrt{s_0}-M_{B}= 0.6\pm0.1\,\rm{GeV}$, the $s_0$ are large enough to take  account of all the ground state contributions sufficiently  but small enough to suppress contaminations from the first radial excited states efficaciously   \cite{WZG-Omega-1P,WZG-Omegab,Wang-2625-2815,WZG-D-wave-lambda,YGL-1D-CPC,WZG-2S-CPC,WZG-Negative-P}.
From Tables \ref{mass}-\ref{mass-GG}, we can see explicitly that for the central values of the baryon masses, the currents with partial derivatives and with covariant derivatives only make tiny differences, while for the central values of the pole residues, the currents with covariant derivatives lead to (slightly) larger values. In general, the  currents with covariant derivatives lead to (slightly) smaller uncertainties comparing to the currents with partial derivatives. Therefore,  we obtain the conclusion tentatively that the currents with the covariant derivatives are better, as the vecuum condensates make slightly larger contributions (see Tables \ref{Borel}-\ref{Borel-GG}) and therefore better QCD sum rules.

Moreover, from Tables \ref{mass}-\ref{mass-GG}, we can see clearly that the uncertainties of the masses are rather small, as we obtain the baryon masses through a fraction, see Eq.\eqref{SR-mass}, the uncertainties originate from the uncertainty  of a parameter in numerator and denominator  are canceled out with each other to a large extent, so the net uncertainties of the masses are small. We can examine the present  calculations by the experimental data in the future.    On the other hand, such a cancellation does not occur for the pole residues, see Eq.\eqref{QCDSR}, and the uncertainties of the pole residues can be as large as $40\%$. In previous works \cite{WZG-Omega-1P,WZG-Omegab,Wang-2625-2815,WZG-D-wave-lambda,YGL-1D-CPC,WZG-Negative-P}, we have given several successful (or reasonable) descriptions of the P-wave and D-wave heavy baryon states with the same traditional  error analysis. In those works, we chose the currents with the partial derivatives, after they were published, we rechecked the calculations by taking the covariant derivatives in stead of the partial derivatives, the predicted baryon masses were as before, while the pole residues were improved slightly, just like in the present work.

 The predicted masses $M=3.35^{+0.13}_{-0.18}\,\rm{GeV}$ ($3.33^{+0.13}_{-0.16}\,\rm{GeV}$), $3.34^{+0.14}_{-0.18}\,\rm{GeV}$ ($3.35^{+0.13}_{-0.16}\,\rm{GeV}$) and $3.35^{+0.12}_{-0.13}\,\rm{GeV}$ ($3.35^{+0.12}_{-0.14}\,\rm{GeV}$) for the $\Omega_c(0,2,{\frac{1}{2}}^+)$,
$\Omega_c(0,2,{\frac{3}{2}}^+)$ and $\Omega_c(0,2,{\frac{5}{2}}^+)$ states for the currents with partial  derivatives (covariant derivatives) are in excellent agreement  with the experimental data $ 3327.1\pm1.2 \mbox{ MeV}$ from the LHCb collaboration \cite{Omegac-Five-LHCb-New}, and support assigning the $\Omega_c(3327)$ to be the $\Sigma$-type D-wave $\Omega_c$ state with the spin-parity $J^P={\frac{1}{2}}^+$, ${\frac{3}{2}}^+$ or ${\frac{5}{2}}^+$. Other predictions can be confronted to the experimental data in the future to diagnose the nature of the D-wave charmed baryon states.

\begin{figure}
 \centering
 \includegraphics[totalheight=5cm,width=7cm]{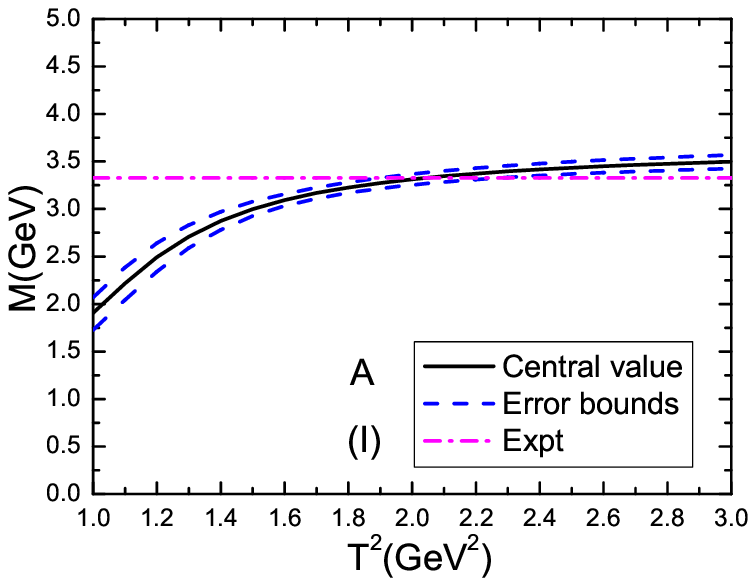}
 \includegraphics[totalheight=5cm,width=7cm]{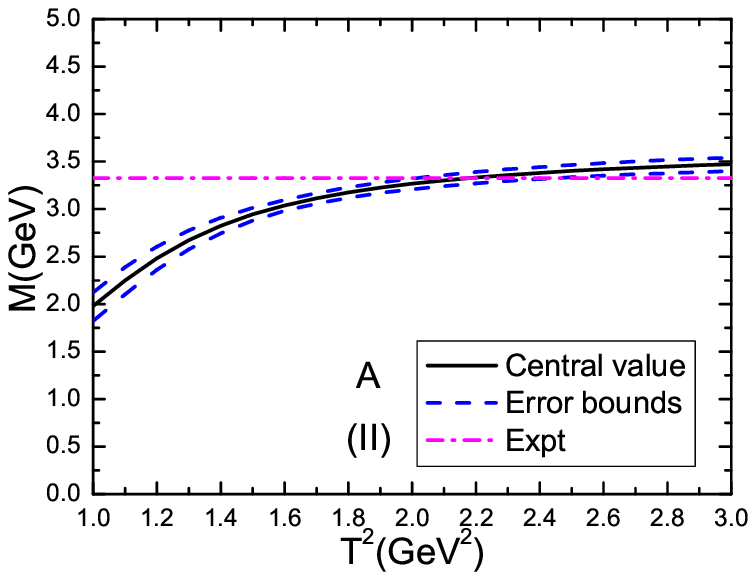}
 \includegraphics[totalheight=5cm,width=7cm]{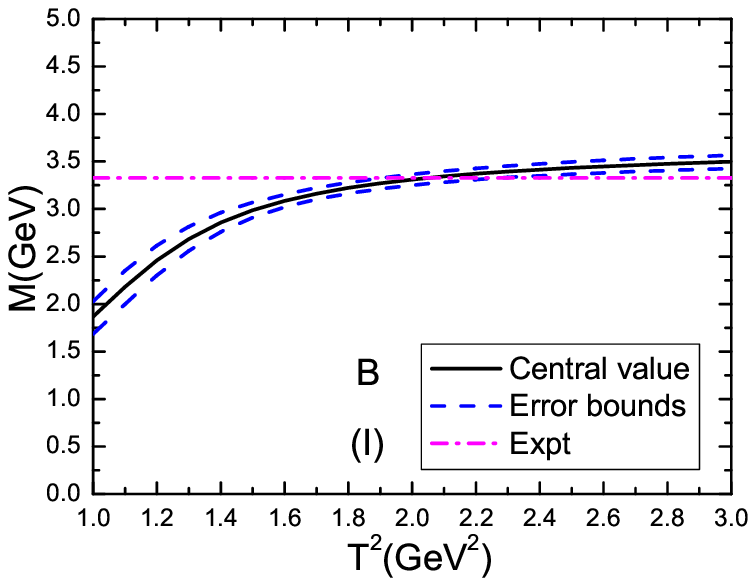}
 \includegraphics[totalheight=5cm,width=7cm]{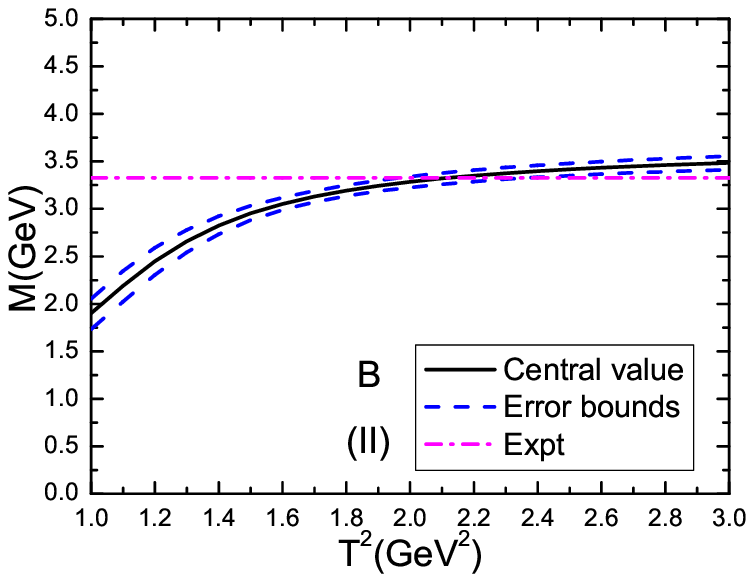}
 \includegraphics[totalheight=5cm,width=7cm]{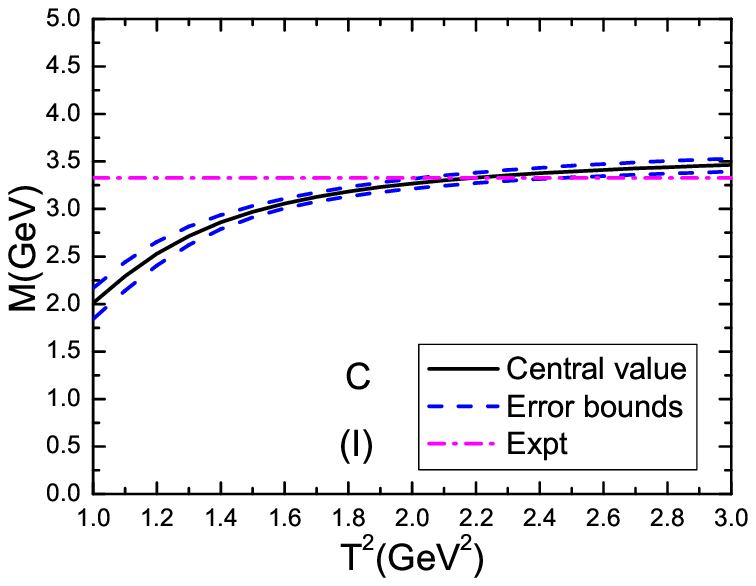}
 \includegraphics[totalheight=5cm,width=7cm]{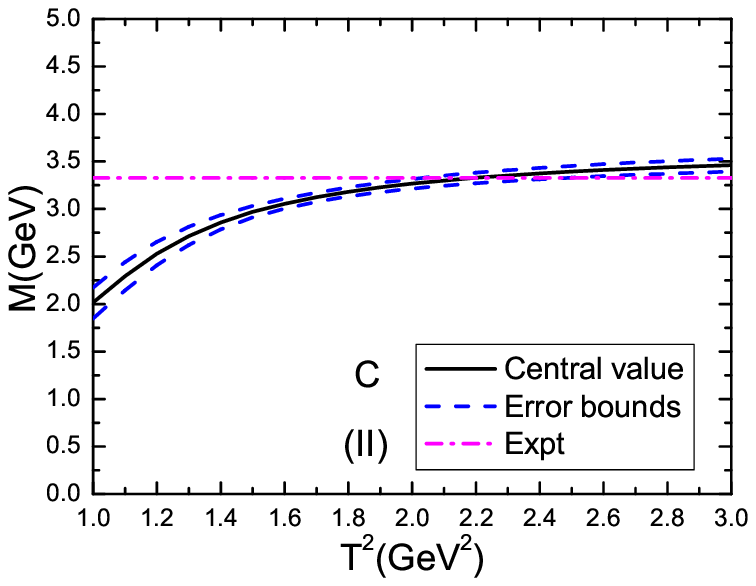}
        \caption{ The masses  of the $\Omega_c(0,2,{\frac{1}{2}}^+)$ ($A$),
$\Omega_c(0,2,{\frac{3}{2}}^+)$ ($B$) and $\Omega_c(0,2,{\frac{5}{2}}^+)$ ($C$) states with variations of the Borel parameters $T^2$, where the (I) and (II) come from the currents with partial derivatives and covariant derivatives, respectively,   the Expt denotes the experimental value of the mass of the $\Omega_c(3327)$.  }\label{Omega-Borel}
\end{figure}

 As an example, in Fig.\ref{Omega-Borel}, we plot the predicted masses of the $\Omega_c(0,2,{\frac{1}{2}}^+)$,
$\Omega_c(0,2,{\frac{3}{2}}^+)$ and $\Omega_c(0,2,{\frac{5}{2}}^+)$ states with variations  of the Borel parameter $T^2$ for the currents with partial  derivatives and covariant derivatives, respectively. From the figure, we can see clearly that the predicted masses increase quickly or slowly with increase
of the Borel parameters, in the Borel windows, the platforms are not flat enough. This maybe due to the fact that
the perturbative contributions dominate the QCD sum rules and the vacuum condensates play a minor role, see Tables \ref{Borel}-\ref{Borel-GG}.

In Ref.\cite{ChenHX-Omega-D-wave}, Mao et al study the D-wave $\Sigma$-type charmed baryon states with the QCD sum rules combined with the heavy quark effective theory, and obtain the predictions $M=3.29^{+0.17}_{-0.25}\,\rm{GeV}$, $3.29^{+0.16}_{-0.25}\,\rm{GeV}$ and $3.49^{+0.30}_{-0.19}\,\rm{GeV}$ for the  $\Omega_c(2,0,{\frac{1}{2}}^+)$,
$\Omega_c(2,0,{\frac{3}{2}}^+)$ and $\Omega_c(0,2,{\frac{5}{2}}^+)$ states,  respectively, which differ from the present predictions
   $3.53^{+0.13}_{-0.17}\,\rm{GeV}$ ($3.52^{+0.12}_{-0.13}\,\rm{GeV}$), $3.54^{+0.12}_{-0.16}\,\rm{GeV}$ ($3.52^{+0.12}_{-0.14}\,\rm{GeV}$) and $3.35^{+0.12}_{-0.13}\,\rm{GeV}$ ($3.35^{+0.12}_{-0.14}\,\rm{GeV}$) significantly, this maybe due to the different Borel windows to extract the baryon masses.

\begin{table}
\begin{center}
\begin{tabular}{|c|c|c|c|c|c|} \hline\hline
$B(L_\rho,L_\lambda,J^P)$        &$T^2(\rm{GeV}^2)$ &$\sqrt{s_0}(\rm{GeV})$ &pole        &perturbative \\ \hline

$\Sigma_c(0,2,{\frac{1}{2}}^+)$  &$1.5-2.1$         &$3.70\pm0.10$          &$(44-86)\%$ &$(83-98)\%$ \\ \hline

$\Sigma_c(0,2,{\frac{3}{2}}^+)$  &$1.5-2.1$         &$3.70\pm0.10$          &$(44-87)\%$ &$(82-97)\%$ \\ \hline

$\Sigma_c(0,2,{\frac{5}{2}}^+)$  &$1.6-2.2$         &$3.70\pm0.10$          &$(46-85)\%$ &$(86-95)\%$ \\ \hline

$\Xi_c(0,2,{\frac{1}{2}}^+)$     &$1.6-2.2$         &$3.80\pm0.10$          &$(46-85)\%$ &$(90-98)\%$ \\ \hline

$\Xi_c(0,2,{\frac{3}{2}}^+)$     &$1.6-2.2$         &$3.80\pm0.10$          &$(46-85)\%$ &$(90-98)\%$ \\ \hline

$\Xi_c(0,2,{\frac{5}{2}}^+)$     &$1.7-2.3$         &$3.80\pm0.10$          &$(47-84)\%$ &$(90-96)\%$ \\ \hline

$\Omega_c(0,2,{\frac{1}{2}}^+)$  &$1.8-2.4$         &$3.95\pm0.10$          &$(46-82)\%$ &$(95-99)\%$ \\ \hline

$\Omega_c(0,2,{\frac{3}{2}}^+)$  &$1.8-2.4$         &$3.95\pm0.10$          &$(46-82)\%$ &$(96-99)\%$ \\ \hline

$\Omega_c(0,2,{\frac{5}{2}}^+)$  &$2.0-2.6$         &$3.95\pm0.10$          &$(44-78)\%$ &$(97-99)\%$ \\ \hline

$\Sigma_c(2,0,{\frac{1}{2}}^+)$  &$1.6-2.2$         &$3.90\pm0.10$          &$(47-86)\%$ &$(122-124)\%$ \\ \hline

$\Sigma_c(2,0,{\frac{3}{2}}^+)$  &$1.7-2.3$         &$3.90\pm0.10$          &$(44-82)\%$ &$(117-119)\%$ \\ \hline

$\Sigma_c(2,0,{\frac{5}{2}}^+)$  &$1.8-2.4$         &$3.90\pm0.10$          &$(44-81)\%$ &$(98-99)\%$ \\ \hline

$\Xi_c(2,0,{\frac{1}{2}}^+)$     &$1.8-2.4$         &$4.00\pm0.10$          &$(47-83)\%$ &$(114-115)\%$ \\ \hline

$\Xi_c(2,0,{\frac{3}{2}}^+)$     &$1.8-2.4$         &$4.00\pm0.10$          &$(47-83)\%$ &$(113-114)\%$ \\ \hline

$\Xi_c(2,0,{\frac{5}{2}}^+)$     &$1.9-2.5$         &$4.00\pm0.10$          &$(46-81)\%$ &$(\sim99)\%$ \\ \hline

$\Omega_c(2,0,{\frac{1}{2}}^+)$  &$2.0-2.6$         &$4.15\pm0.10$          &$(48-82)\%$ &$(109-110)\%$ \\ \hline

$\Omega_c(2,0,{\frac{3}{2}}^+)$  &$2.0-2.6$         &$4.15\pm0.10$          &$(49-82)\%$ &$(110-111)\%$ \\ \hline

$\Omega_c(2,0,{\frac{5}{2}}^+)$  &$2.1-2.7$         &$4.15\pm0.10$          &$(47-79)\%$ &$(\sim99)\%$ \\ \hline

\end{tabular}
\end{center}
\caption{ The Borel parameters $T^2$, continuum threshold parameters $s_0$, pole contributions (from the ground states) and perturbative contributions for the D-wave (with partial derivatives) charmed baryon states.}\label{Borel}
\end{table}

\begin{table}
\begin{center}
\begin{tabular}{|c|c|c|c|c|c|} \hline\hline
$B(L_\rho,L_\lambda,J^P)$        &$T^2(\rm{GeV}^2)$ &$\sqrt{s_0}(\rm{GeV})$ &pole        &perturbative \\ \hline

$\Sigma_c(0,2,{\frac{1}{2}}^+)$  &$1.6-2.2$         &$3.70\pm0.10$          &$(42-83)\%$ &$(76-87)\%$ \\ \hline

$\Sigma_c(0,2,{\frac{3}{2}}^+)$  &$1.6-2.2$         &$3.70\pm0.10$          &$(41-82)\%$ &$(82-94)\%$ \\ \hline

$\Sigma_c(0,2,{\frac{5}{2}}^+)$  &$1.7-2.3$         &$3.70\pm0.10$          &$(42-80)\%$ &$(88-95)\%$ \\ \hline

$\Xi_c(0,2,{\frac{1}{2}}^+)$     &$1.7-2.3$         &$3.80\pm0.10$          &$(44-82)\%$ &$(80-88)\%$ \\ \hline

$\Xi_c(0,2,{\frac{3}{2}}^+)$     &$1.7-2.3$         &$3.80\pm0.10$          &$(43-82)\%$ &$(87-94)\%$ \\ \hline

$\Xi_c(0,2,{\frac{5}{2}}^+)$     &$1.8-2.4$         &$3.80\pm0.10$          &$(44-80)\%$ &$(91-96)\%$ \\ \hline

$\Omega_c(0,2,{\frac{1}{2}}^+)$  &$1.9-2.5$         &$3.95\pm0.10$          &$(44-80)\%$ &$(84-90)\%$ \\ \hline

$\Omega_c(0,2,{\frac{3}{2}}^+)$  &$1.9-2.5$         &$3.95\pm0.10$          &$(44-79)\%$ &$(91-96)\%$ \\ \hline

$\Omega_c(0,2,{\frac{5}{2}}^+)$  &$2.0-2.6$         &$3.95\pm0.10$          &$(44-78)\%$ &$(96-98)\%$ \\ \hline

$\Sigma_c(2,0,{\frac{1}{2}}^+)$  &$1.9-2.5$         &$3.90\pm0.10$          &$(44-79)\%$ &$(79-86)\%$ \\ \hline

$\Sigma_c(2,0,{\frac{3}{2}}^+)$  &$2.0-2.6$         &$3.90\pm0.10$          &$(40-75)\%$ &$(81-86)\%$ \\ \hline

$\Sigma_c(2,0,{\frac{5}{2}}^+)$  &$1.8-2.4$         &$3.90\pm0.10$          &$(44-81)\%$ &$(98-99)\%$ \\ \hline

$\Xi_c(2,0,{\frac{1}{2}}^+)$     &$2.1-2.7$         &$4.00\pm0.10$          &$(42-75)\%$ &$(82-87)\%$ \\ \hline

$\Xi_c(2,0,{\frac{3}{2}}^+)$     &$2.1-2.7$         &$4.00\pm0.10$          &$(42-76)\%$ &$(83-87)\%$ \\ \hline

$\Xi_c(2,0,{\frac{5}{2}}^+)$     &$1.9-2.5$         &$4.00\pm0.10$          &$(46-81)\%$ &$(\sim99)\%$ \\ \hline

$\Omega_c(2,0,{\frac{1}{2}}^+)$  &$2.3-2.9$         &$4.15\pm0.10$          &$(43-74)\%$ &$(85-89)\%$ \\ \hline

$\Omega_c(2,0,{\frac{3}{2}}^+)$  &$2.3-2.9$         &$4.15\pm0.10$          &$(43-75)\%$ &$(86-89)\%$ \\ \hline

$\Omega_c(2,0,{\frac{5}{2}}^+)$  &$2.1-2.7$         &$4.15\pm0.10$          &$(47-80)\%$ &$(\sim99)\%$ \\ \hline

\end{tabular}
\end{center}
\caption{ The Borel parameters $T^2$, continuum threshold parameters $s_0$, pole contributions (from the ground states) and perturbative contributions for the D-wave (with covariant derivatives) charmed baryon states.}\label{Borel-GG}
\end{table}

\begin{table}
\begin{center}
\begin{tabular}{|c|c|c|c|c|c|c|}\hline\hline
$B(L_\rho,L_\lambda,J^P)$       &$M(\rm{GeV})$   &$\lambda(10^{-1}\rm{GeV}^5)$    &assignments   \\\hline

$\Sigma_c(0,2,{\frac{1}{2}}^+)$ &$3.07^{+0.18}_{-0.34}$  &$1.25^{+0.46}_{-0.55}$  &   \\ \hline

$\Sigma_c(0,2,{\frac{3}{2}}^+)$ &$3.05^{+0.19}_{-0.35}$  &$0.68^{+0.25}_{-0.31}$  &   \\ \hline

$\Sigma_c(0,2,{\frac{5}{2}}^+)$ &$3.08^{+0.15}_{-0.23}$  &$0.95^{+0.29}_{-0.31}$  &   \\ \hline

$\Xi_c(0,2,{\frac{1}{2}}^+)$    &$3.19^{+0.15}_{-0.25}$  &$1.56^{+0.49}_{-0.55}$  &   \\ \hline

$\Xi_c(0,2,{\frac{3}{2}}^+)$    &$3.18^{+0.16}_{-0.26}$  &$0.85^{+0.27}_{-0.31}$  &   \\ \hline

$\Xi_c(0,2,{\frac{5}{2}}^+)$    &$3.18^{+0.14}_{-0.19}$  &$1.15^{+0.30}_{-0.32}$  &   \\ \hline

$\Omega_c(0,2,{\frac{1}{2}}^+)$ &$3.35^{+0.13}_{-0.18}$  &$2.98^{+0.77}_{-0.79}$  &$\Omega_c(3327)$   \\ \hline

$\Omega_c(0,2,{\frac{3}{2}}^+)$ &$3.34^{+0.14}_{-0.18}$  &$1.63^{+0.42}_{-0.44}$  &$\Omega_c(3327)$    \\ \hline

$\Omega_c(0,2,{\frac{5}{2}}^+)$ &$3.35^{+0.12}_{-0.13}$  &$2.17^{+0.47}_{-0.45}$  &$\Omega_c(3327)$    \\ \hline

$\Sigma_c(2,0,{\frac{1}{2}}^+)$ &$3.33^{+0.15}_{-0.30}$  &$1.84^{+0.58}_{-0.74}$  &   \\ \hline

$\Sigma_c(2,0,{\frac{3}{2}}^+)$ &$3.34^{+0.14}_{-0.24}$  &$1.06^{+0.30}_{-0.36}$  &   \\ \hline

$\Sigma_c(2,0,{\frac{5}{2}}^+)$ &$3.33^{+0.12}_{-0.13}$  &$3.33^{+0.77}_{-0.74}$  &   \\ \hline

$\Xi_c(2,0,{\frac{1}{2}}^+)$    &$3.41^{+0.14}_{-0.20}$  &$2.26^{+0.59}_{-0.64}$  &   \\ \hline

$\Xi_c(2,0,{\frac{3}{2}}^+)$    &$3.41^{+0.13}_{-0.20}$  &$1.25^{+0.32}_{-0.36}$  &   \\ \hline

$\Xi_c(2,0,{\frac{5}{2}}^+)$    &$3.41^{+0.12}_{-0.13}$  &$3.88^{+0.86}_{-0.82}$  &   \\ \hline

$\Omega_c(2,0,{\frac{1}{2}}^+)$ &$3.53^{+0.13}_{-0.17}$  &$4.10^{+0.95}_{-0.96}$  &   \\ \hline

$\Omega_c(2,0,{\frac{3}{2}}^+)$ &$3.54^{+0.12}_{-0.16}$  &$2.26^{+0.52}_{-0.51}$  &   \\ \hline

$\Omega_c(2,0,{\frac{5}{2}}^+)$ &$3.54^{+0.11}_{-0.13}$  &$6.92^{+1.43}_{-1.33}$  &   \\ \hline

      \hline
\end{tabular}
\end{center}
\caption{ The masses  and pole residues of the D-wave (with partial derivatives) charmed baryon states.} \label{mass}
\end{table}

\begin{table}
\begin{center}
\begin{tabular}{|c|c|c|c|c|c|c|}\hline\hline
$B(L_\rho,L_\lambda,J^P)$       &$M(\rm{GeV})$   &$\lambda(10^{-1}\rm{GeV}^5)$    &assignments   \\\hline

$\Sigma_c(0,2,{\frac{1}{2}}^+)$ &$3.08^{+0.17}_{-0.25}$  &$1.37^{+0.43}_{-0.48}$  &   \\ \hline

$\Sigma_c(0,2,{\frac{3}{2}}^+)$ &$3.09^{+0.17}_{-0.28}$  &$0.73^{+0.24}_{-0.27}$  &   \\ \hline

$\Sigma_c(0,2,{\frac{5}{2}}^+)$ &$3.11^{+0.15}_{-0.19}$  &$1.01^{+0.27}_{-0.29}$  &   \\ \hline

$\Xi_c(0,2,{\frac{1}{2}}^+)$    &$3.18^{+0.15}_{-0.20}$  &$1.66^{+0.47}_{-0.49}$  &   \\ \hline

$\Xi_c(0,2,{\frac{3}{2}}^+)$    &$3.20^{+0.14}_{-0.22}$  &$0.89^{+0.26}_{-0.27}$  &   \\ \hline

$\Xi_c(0,2,{\frac{5}{2}}^+)$    &$3.21^{+0.13}_{-0.17}$  &$1.20^{+0.29}_{-0.30}$  &   \\ \hline

$\Omega_c(0,2,{\frac{1}{2}}^+)$ &$3.33^{+0.13}_{-0.16}$  &$3.10^{+0.76}_{-0.75}$  &$\Omega_c(3327)$   \\ \hline

$\Omega_c(0,2,{\frac{3}{2}}^+)$ &$3.35^{+0.13}_{-0.16}$  &$1.68^{+0.41}_{-0.41}$  &$\Omega_c(3327)$    \\ \hline

$\Omega_c(0,2,{\frac{5}{2}}^+)$ &$3.35^{+0.12}_{-0.14}$  &$2.18^{+0.47}_{-0.46}$  &$\Omega_c(3327)$    \\ \hline

$\Sigma_c(2,0,{\frac{1}{2}}^+)$ &$3.28^{+0.13}_{-0.17}$  &$2.11^{+0.53}_{-0.51}$  &   \\ \hline

$\Sigma_c(2,0,{\frac{3}{2}}^+)$ &$3.30^{+0.13}_{-0.16}$  &$1.20^{+0.27}_{-0.28}$  &   \\ \hline

$\Sigma_c(2,0,{\frac{5}{2}}^+)$ &$3.33^{+0.12}_{-0.13}$  &$3.33^{+0.77}_{-0.74}$  &   \\ \hline

$\Xi_c(2,0,{\frac{1}{2}}^+)$    &$3.39^{+0.12}_{-0.15}$  &$2.56^{+0.56}_{-0.54}$  &   \\ \hline

$\Xi_c(2,0,{\frac{3}{2}}^+)$    &$3.38^{+0.13}_{-0.14}$  &$1.40^{+0.31}_{-0.30}$  &   \\ \hline

$\Xi_c(2,0,{\frac{5}{2}}^+)$    &$3.41^{+0.12}_{-0.13}$  &$3.88^{+0.86}_{-0.82}$  &   \\ \hline

$\Omega_c(2,0,{\frac{1}{2}}^+)$ &$3.52^{+0.12}_{-0.13}$  &$4.57^{+0.92}_{-0.87}$  &   \\ \hline

$\Omega_c(2,0,{\frac{3}{2}}^+)$ &$3.52^{+0.12}_{-0.14}$  &$2.49^{+0.51}_{-0.47}$  &   \\ \hline

$\Omega_c(2,0,{\frac{5}{2}}^+)$ &$3.54^{+0.11}_{-0.13}$  &$6.92^{+1.43}_{-1.33}$  &   \\ \hline
      \hline
\end{tabular}
\end{center}
\caption{ The masses  and pole residues of the D-wave (with covariant derivatives) charmed baryon states.} \label{mass-GG}
\end{table}

\section{Conclusion}
In this article, we construct the $\Sigma$-type interpolating currents with both the partial derivatives and covariant derivatives to explore  the D-wave charmed baryon states via
the QCD sum rules in a systematic way. We carry out the operator product expansion up to the vacuum condensates of dimension $10$  consistently, and distinguish  the contributions of the positive and negative parity baryon states unambiguously,
 then  we investigate  the masses and pole residues of the ground states in details,  the  predicted masses $M=3.35^{+0.13}_{-0.18}\,\rm{GeV}$ ($3.33^{+0.13}_{-0.16}\,\rm{GeV}$), $3.34^{+0.14}_{-0.18}\,\rm{GeV}$ ($3.35^{+0.13}_{-0.16}\,\rm{GeV}$) and $3.35^{+0.12}_{-0.13}\,\rm{GeV}$ ($3.35^{+0.12}_{-0.14}\,\rm{GeV}$) for the $\Omega_c(0,2,{\frac{1}{2}}^+)$,
$\Omega_c(0,2,{\frac{3}{2}}^+)$ and $\Omega_c(0,2,{\frac{5}{2}}^+)$ states from the currents with the partial derivatives (covariant derivatives) are in excellent agreement  with the experimental data $ 3327.1\pm1.2 \mbox{ MeV}$ from the LHCb collaboration, and support assigning the $\Omega_c(3327)$ to be the $\Sigma$-type D-wave $\Omega_c$ state with the spin-parity $J^P={\frac{1}{2}}^+$, ${\frac{3}{2}}^+$ or ${\frac{5}{2}}^+$. Other predictions can be confronted to the experimental data in the future to diagnose the nature of the D-wave charmed baryon states.
In general, the  currents with covariant derivatives lead to (slightly) smaller uncertainties comparing to the currents with partial derivatives, and we prefer the  covariant derivatives in constructing the interpolating currents.

\section*{Appendix}
The QCD spectral densities for the currents with the partial derivatives,
\begin{eqnarray}
\rho_{j,QCD}^1(s)&=&\rho_{j,\Sigma_c}^1(s)\, ,\,\, \rho_{j,\Xi_c}^1(s)\, , \, \,\rho_{j,\Omega_c}^1(s)\, , \nonumber \\
\rho_{j,QCD}^0(s)&=&m_c\rho_{j,\Sigma_c}^0(s)\, ,\,\, m_c\rho_{j,\Xi_c}^0(s)\, , \, \,m_c\rho_{j,\Omega_c}^0(s)\, ,
\end{eqnarray}
where $j=\frac{1}{2}$, $\frac{3}{2}$, $\frac{5}{2}$,
\begin{eqnarray}
\rho_{j,\Sigma_c}^1(s)&=&\rho_{j,\Xi_c}^1(s)\mid_{m_s \to 0, \langle\bar{s}s\rangle \to\langle\bar{q}q\rangle, \langle\bar{s}g_s\sigma Gs\rangle \to\langle\bar{q}g_s\sigma Gq\rangle}\, ,
\end{eqnarray}

\begin{eqnarray}
\rho_{j,\Sigma_c}^0(s)&=&\rho_{j,\Xi_c}^0(s)\mid_{m_s \to 0, \langle\bar{s}s\rangle \to\langle\bar{q}q\rangle, \langle\bar{s}g_s\sigma Gs\rangle \to\langle\bar{q}g_s\sigma Gq\rangle}\, ,
\end{eqnarray}

\begin{eqnarray}
\rho^1_{\frac{1}{2},\Xi_c}(s)&=&\frac{1} {128\pi^4} \int_{x_i}^{1}dxx(1-x)^4(s-\tilde{m}_c^2)^4\nonumber\\
&&-\frac{24(1-r)m_s\langle\bar{q}g_s\sigma Gq\rangle+(4+3r)m_s\langle\bar{s}g_s\sigma Gs\rangle}{32\pi^2}\int_{x_i}^{1}dx\,x(1-x)(s-\tilde{m}_c^2)\nonumber\\
&&-\frac{m^2_c}{96\pi^2}\langle\frac{\alpha_{s}GG}{\pi}\rangle\int_{x_i}^{1}dx\frac{(1-x)^4}{x^2}(s-\tilde{m}_c^2)\nonumber\\
&&+\frac{1}{256\pi^2}\langle\frac{\alpha_{s}GG}{\pi}\rangle\int_{x_i}^{1}dx\,(1-x)^2(16rx-19x+3r)
(s-\tilde{m}_c^2)^2\nonumber\\
&&+\frac{(3r-4)m_s\langle\bar{s}s\rangle}{192}\langle\frac{\alpha_{s}GG}{\pi}\rangle\int_{x_i}^{1}dx(1-x)\nonumber\\
&&+\frac{3\langle\bar{q}g_s\sigma Gq\rangle\langle\bar{s}g_s\sigma Gs\rangle}{16}\delta(s-m_c^2)\, ,
\end{eqnarray}

\begin{eqnarray}
\rho^0_{\frac{1}{2},\Xi_c}(s)&=&\frac{3}{512\pi^4}\int_{x_i}^{1}dx(1-x)^4(s-\tilde{m}_c^2)^4
-\frac{m_c^2}{128\pi^2}\langle\frac{\alpha_{s}GG}{\pi}\rangle\int_{x_i}^{1}dx\frac{(1-x)^4}{x^3}(s-\tilde{m}_c^2)\nonumber\\
&&+\frac{3}{256\pi^2}\langle\frac{\alpha_{s}GG}{\pi}\rangle\int_{x_i}^{1}
dx\frac{(1-x)^4}{x^2}(s-\tilde{m}_c^2)^2\nonumber\\
&&+\frac{1}{768\pi^2}\langle\frac{\alpha_{s}GG}{\pi}\rangle\int_{x_i}^{1}dx\frac{(1-x)^2
[-4-35x+r(-2+41x)]}{x}(s-\tilde{m}_c^2)^2 \nonumber\\
&&+\frac{3\langle\bar{q}g_s\sigma Gq\rangle\langle\bar{s}g_s\sigma Gs\rangle}{8}\delta(s-m_c^2)\, ,
\end{eqnarray}

\begin{eqnarray}
\rho^1_{\frac{3}{2},\Xi_c}(s)&=&\frac{1} {3072\pi^4} \int_{x_i}^{1}dx\, x(x+6)(1-x)^4(s-\tilde{m}_c^2)^4\nonumber\\
&&-\frac{(4+3r)m_s\langle\bar{s}g_s\sigma Gs\rangle+24(1-r)m_s\langle\bar{q}g_s\sigma Gq\rangle}{192\pi^2}\int_{x_i}^{1}dx\,x^2(1-x)(s-\tilde{m}_c^2)\nonumber\\
&&-\frac{m_c^2}{2304\pi^2}\langle\frac{\alpha_{s}GG}{\pi}\rangle \int_{x_i}^{1}dx\frac{(1-x)^4(6+x)}{x^2}(s-\tilde{m}_c^2)\nonumber\\
&&+\frac{r-1}{4608\pi^2}\langle\frac{\alpha_{s}GG}{\pi}\rangle \int_{x_i}^{1}dx\,x(1-x)^2(71+25x)(s-\tilde{m}_c^2)^2\nonumber\\
&&+\frac{3\langle\bar{q}g_s\sigma Gq\rangle\langle\bar{s}g_s\sigma Gs\rangle}{32}\delta(s-m_c^2)\, ,
\end{eqnarray}

\begin{eqnarray}
\rho^0_{\frac{3}{2},\Xi_c}(s)&=&\frac{1} {3072\pi^4} \int_{x_i}^{1}dx\,(x+6)(1-x)^4(s-\tilde{m}_c^2)^4\nonumber\\
&&-\frac{(4+3r)m_s\langle\bar{s}g_s\sigma Gs\rangle+24(1-r)m_s\langle\bar{q}g_s\sigma Gq\rangle}{192\pi^2}\int_{x_i}^{1}dx\,x(1-x)(s-\tilde{m}_c^2)\nonumber\\
&&-\frac{m_c^2}{2304\pi^2}\langle\frac{\alpha_{s}GG}{\pi}\rangle \int_{x_i}^{1}dx\frac{(1-x)^4(6+x)}{x^3}(s-\tilde{m}_c^2)\nonumber\\
&&+\frac{1}{1536\pi^2}\langle\frac{\alpha_{s}GG}{\pi}\rangle \int_{x_i}^{1}dx\frac{(1-x)^4(6+x)}{x^2}(s-\tilde{m}_c^2)^2\nonumber\\
&&+\frac{r-1}{4608\pi^2}\langle\frac{\alpha_{s}GG}{\pi}\rangle \int_{x_i}^{1}dx\,(1-x)^2(71+25x)(s-\tilde{m}_c^2)^2 \nonumber\\
&&+\frac{3\langle\bar{q}g_s\sigma Gq\rangle\langle\bar{s}g_s\sigma Gs\rangle}{32}\delta(s-m_c^2)\, ,
\end{eqnarray}

\begin{eqnarray}
\rho^1_{\frac{5}{2},\Xi_c}(s)&=&\frac{1} {960\pi^4} \int_{x_i}^{1}dx\,x(1-x)^4[8+3x+14x^2+r(-8-3x+11x^2)](s-\tilde{m}_c^2)^4\nonumber\\
&&+\frac{m_s\langle\bar{s}s\rangle}{4\pi^2}\int_{x_i}^{1}dx\,x^2(1-x)^2(-1+r+4x-rx)(s-\tilde{m}_c^2)^2\nonumber\\
&&-\frac{m_s\langle\bar{q}q\rangle}{4\pi^2}\int_{x_i}^{1}dx\,x^3(1-x)^2(s-\tilde{m}_c^2)^2\nonumber\\
&&+\frac{m_s\langle\bar{s}g_s\sigma Gs\rangle}{12\pi^2}\int_{x_i}^{1}dx\, x(1-x)[-4+21x-19x^2+r(4-21x+17x^2)](s-\tilde{m}_c^2)\nonumber\\
&&-\frac{8(1-r)m_s\langle\bar{q}g_s\sigma Gq\rangle+m_s\langle\bar{q}g_s\sigma Gq\rangle}{16\pi^2}\int_{x_i}^{1}dx\,x^2(1-x)^2(s-\tilde{m}_c^2)\nonumber\\
&&-\frac{m_c^2}{720\pi^2}\langle\frac{\alpha_{s}GG}{\pi}\rangle \int_{x_i}^{1}dx\frac{(1-x)^4[8+3x+14x^2+r(-8-3x+11x^2)]}{x^2}(s-\tilde{m}_c^2)\nonumber\\
&&+\frac{1}{480\pi^2}\langle\frac{\alpha_{s}GG}{\pi}\rangle\int_{x_i}^{1}dx\,
(1-x)^3[8+4x+18x^2+r(-8-4x+7x^2)](s-\tilde{m}_c^2)^2\nonumber\\
&&+\frac{1}{288\pi^2}\langle\frac{\alpha_{s}GG}{\pi}\rangle\int_{x_i}^{1}dx\,
x(1-x)^2[-2-x+3x^2+r(2+x+3x^2)](s-\tilde{m}_c^2)^2
\nonumber\\
&&+\frac{m_sm_c^2\langle\bar{s}s\rangle}{36} \langle\frac{\alpha_{s}GG}{\pi}\rangle\int_{x_i}^{1}dx\frac{(1-x)^2[1+r(x-1)-4x]}{x}\delta(s-\tilde{m}_c^2)\nonumber\\
&&+\frac{m_sm_c^2\langle\bar{q}q\rangle}{36} \langle\frac{\alpha_{s}GG}{\pi}\rangle\int_{x_i}^{1}dx(1-x)^2\delta(s-\tilde{m}_c^2)\nonumber\\
&&-\frac{(3-r)m_s\langle\bar{s}s\rangle}{72}\langle\frac{\alpha_{s}GG}{\pi}\rangle \int_{x_i}^{1}dx(x-1)x^2\nonumber\\
&&+\frac{\langle\bar{q}g_s\sigma Gq\rangle\langle\bar{s}g_s\sigma Gs\rangle}{12}\delta(s-m_c^2)\, ,
\end{eqnarray}

\begin{eqnarray}
\rho^0_{\frac{5}{2},\Xi_c}(s)&=&\frac{1} {1920\pi^4}\int_{x_i}^{1}dx\,(1-x)^4[12+4x+9x^2+2r(-6-2x+3x^2)](s-\tilde{m}_c^2)^4\nonumber\\
&&-\frac{(2r-3)m_s\langle\bar{s}s\rangle+4m_s\langle\bar{q}q\rangle}{8\pi^2}\int_{x_i}^{1}dx\,
x^2(1-x)^2(s-\tilde{m}_c^2)^2\nonumber\\
&&-\frac{(1-r)m_s\langle\bar{s}g_s\sigma Gs\rangle}{12\pi^2}\int_{x_i}^{1}dxx(1-x)(-8+9x)(s-\tilde{m}_c^2)\nonumber\\
&&+\frac{(8r-9)m_s\langle\bar{q}g_s\sigma Gq\rangle}{8\pi^2}\int_{x_i}^{1}dx\,x(1-x)^2(s-\tilde{m}_c^2)\nonumber\\
&&-\frac{m_c^2}{1440\pi^2}\langle\frac{\alpha_{s}GG}{\pi}\rangle \int_{x_i}^{1}dx\frac{(1-x)^4[12+4x+9x^2+2r(-6-2x+3x^2)]}{x^3}(s-\tilde{m}_c^2)\nonumber\\
&&+\frac{1}{960\pi^2}\langle\frac{\alpha_{s}GG}{\pi}\rangle\int_{x_i}^{1}dx\frac{(1-x)^4[12+4x+9x^2+2r(-6-2x+3x^2)]
}{x^2}(s-\tilde{m}_c^2)^2\nonumber\\
&&-\frac{1}{160\pi^2}\langle\frac{\alpha_{s}GG}{\pi}\rangle\int_{x_i}^{1}dx\frac{(x-1)^3[r(-4-2x+x^2)+2(2+x+2x^2)]
}{x}(s-\tilde{m}_c^2)^2\nonumber\\
&&-\frac{1}{192\pi^2}\langle\frac{\alpha_{s}GG}{\pi}\rangle\int_{x_i}^{1}dx(x-1)^2[4+2x+3x^2-2r(2+x)]
(s-\tilde{m}_c^2)^2 \nonumber\\
&&+\frac{(2r-3)m_sm_c^2\langle\bar{s}s\rangle}{72} \langle\frac{\alpha_{s}GG}{\pi}\rangle\int_{x_i}^{1}dx\frac{(1-x)^2}{x}\delta(s-\tilde{m}_c^2)\nonumber\\
&&+\frac{m_sm_c^2\langle\bar{q}q\rangle}{18} \langle\frac{\alpha_{s}GG}{\pi}\rangle\int_{x_i}^{1}dx\frac{(1-x)^2}{x}\delta(s-\tilde{m}_c^2)\nonumber\\
&&-\frac{(2r-3)m_s\langle\bar{s}s\rangle}{24}\langle\frac{\alpha_{s}GG}{\pi}\rangle \int_{x_i}^{1}dx(x-1)^2\nonumber\\
&&-\frac{m_s\langle\bar{q}q\rangle}{6}\langle\frac{\alpha_{s}GG}{\pi}\rangle \int_{x_i}^{1}dx\, (x-1)^2+\frac{(r-1)m_s\langle\bar{s}s\rangle}{24}\langle\frac{\alpha_{s}GG}{\pi}\rangle \int_{x_i}^{1}dx\,(x-1)x\nonumber\\
&&+\frac{\langle\bar{q}g_s\sigma Gq\rangle\langle\bar{s}g_s\sigma Gs\rangle}{6}\delta(s-m_c^2)\, ,
\end{eqnarray}

\begin{eqnarray}
\rho_{j,\Omega_c}^1(s)&=&2\rho_{j,\Xi_c}^1(s)\mid_{m_s \to 2m_s, \langle\bar{q}q\rangle \to\langle\bar{s}s\rangle, \langle\bar{q}g_s\sigma Gq\rangle \to\langle\bar{s}g_s\sigma Gs\rangle}\, ,
\end{eqnarray}

\begin{eqnarray}
\rho_{j,\Omega_c}^0(s)&=&2\rho_{j,\Xi_c}^0(s)\mid_{m_s \to 2m_s, \langle\bar{q}q\rangle \to\langle\bar{s}s\rangle, \langle\bar{q}g_s\sigma Gq\rangle \to\langle\bar{s}g_s\sigma Gs\rangle}\, ,
\end{eqnarray}
where $x_i=\frac{m_c^2}{s}$, and $r=+1$ and $-1$ for the currents with the quantum numbers $(L_\rho,L_\lambda)=(0,2)$ and $(2,0)$, respectively.

With the simple replacements,
\begin{eqnarray}
\rho_{j,QCD}^1(s)&\to&\rho_{j,QCD}^1(s)+\rho_{j,QCD}^{A,1}(s)\, , \nonumber \\
\rho_{j,QCD}^0(s)&\to&\rho_{j,QCD}^0(s)+\rho_{j,QCD}^{A,0}(s)\, ,
\end{eqnarray}
we obtain the corresponding QCD spectral densities for the currents with the covariant derivatives,
where the additional ($A$) terms,
\begin{eqnarray}
\rho_{j,QCD}^{A,1}(s)&=&\rho_{j,\Sigma_c}^{A,1}(s)\, ,\,\, \rho_{j,\Xi_c}^{A,1}(s)\, , \, \,\rho_{j,\Omega_c}^{A,1}(s)\, , \nonumber \\
\rho_{j,QCD}^{A,0}(s)&=&m_c\rho_{j,\Sigma_c}^{A,0}(s)\, ,\,\, m_c \rho_{j,\Xi_c}^{A,0}(s)\, , \, \, m_c\rho_{j,\Omega_c}^{A,0}(s)\, ,
\end{eqnarray}

\begin{eqnarray}
\rho_{j,\Sigma_c}^{A,1}(s)&=&\rho_{j,\Xi_c}^{A,1}(s)\mid_{m_s \to 0, \langle\bar{s}s\rangle \to\langle\bar{q}q\rangle, \langle\bar{s}g_s\sigma Gs\rangle \to\langle\bar{q}g_s\sigma Gq\rangle}\, ,
\end{eqnarray}

\begin{eqnarray}
\rho_{j,\Sigma_c}^{A,0}(s)&=&\rho_{j,\Xi_c}^{A,0}(s)\mid_{m_s \to 0, \langle\bar{s}s\rangle \to\langle\bar{q}q\rangle, \langle\bar{s}g_s\sigma Gs\rangle \to\langle\bar{q}g_s\sigma Gq\rangle}\, ,
\end{eqnarray}

\begin{eqnarray}
\rho^{A,1}_{\frac{1}{2},\Xi_c}(s)&=&\frac{5(1-2r)m_s\langle\bar{q}g_s\sigma Gq\rangle}{16\pi^2}\int_{x_i}^{1}dxx(1-x)({s-\tilde{m}_c^2})\nonumber\\
&&+\frac{119-46r}{512\pi^2}\langle\frac{\alpha_{s}GG}{\pi}\rangle\int_{x_i}^{1}dxx(1-x)^2({s-\tilde{m}_c^2})^2\nonumber\\
&&+\frac{(47-46r)m_s\langle\bar{s}s\rangle}{192}\langle\frac{\alpha_{s}GG}{\pi}\rangle\int_{x_i}^{1}dxx\nonumber\\
&& +\frac{(2-r)m_s\langle\bar{q}q\rangle}{48}\langle\frac{\alpha_{s}GG}{\pi}\rangle\int_{x_i}^{1}dxx \,,
\end{eqnarray}

\begin{eqnarray}
\rho^{A,0}_{\frac{1}{2},\Xi_c}(s)&=&
\frac{1}{384\pi^2}\langle\frac{\alpha_{s}GG}{\pi}\rangle\int_{x_i}^{1}dx\frac{(1-x)^2[4-7x+2r(2+x)]}{x}(s-\tilde{m}_c^2)^2\nonumber\\
&&-\frac{(1-2r)m_s\langle\bar{s}s\rangle}{32}\langle\frac{\alpha_{s}GG}{\pi}\rangle\int_{x_i}^{1}dx
-\frac{(2-r)m_s\langle\bar{q}q\rangle}{24}\langle\frac{\alpha_{s}GG}{\pi}\rangle\int_{x_i}^{1}dx\nonumber\\
&&+\frac{6-9r}{64\pi^2}\langle\frac{\alpha_{s}GG}{\pi}\rangle\int_{x_i}^{1}dx(1-x)^2({s-\tilde{m}_c^2})^2\nonumber\\
&&+\frac{(2-r)m_s\langle\bar{s}s\rangle}{48}\langle\frac{\alpha_{s}GG}{\pi}\rangle\int_{x_i}^{1}dx\, ,
\end{eqnarray}

\begin{eqnarray}
\rho^{A,1}_{\frac{3}{2},\Xi_c}(s)&=&-\frac{5(1-2r)m_s\langle\bar{q}g_s\sigma Gq\rangle}{96\pi^2}\int_{x_i}^{1}dxx^2(x-1)({s-\tilde{m}_c^2})\nonumber\\
&&+\frac{163-110r}{3072\pi^2}\langle\frac{\alpha_{s}GG}{\pi}\rangle\int_{x_i}^{1}dxx(1-x)^2({s-\tilde{m}_c^2})^2\nonumber\\
&&+\frac{14r-97}{4608\pi^2}\langle\frac{\alpha_{s}GG}{\pi}\rangle\int_{x_i}^{1}dxx(1-x)^3({s-\tilde{m}_c^2})^2\nonumber\\
&&+\frac{(49-38r)m_s\langle\bar{s}s\rangle}{1152}\langle\frac{\alpha_{s}GG}{\pi}\rangle\int_{x_i}^{1}dxx\nonumber\\
&&+\frac{(2-r)m_s\langle\bar{q}q\rangle}{288}\langle\frac{\alpha_{s}GG}{\pi}\rangle\int_{x_i}^{1}dxx(4x-5)\nonumber\\
&&+\frac{(25r-23)m_s\langle\bar{s}s\rangle}{288}\langle\frac{\alpha_{s}GG}{\pi}\rangle\int_{x_i}^{1}dxx(1-x)\,,
\end{eqnarray}

\begin{eqnarray}
\rho^{A,0}_{\frac{3}{2},\Xi_c}(s)&=&-\frac{5(1-2r)m_s\langle\bar{q}g_s\sigma Gq\rangle}{96\pi^2}\int_{x_i}^{1}dxx(x-1)({s-\tilde{m}_c^2})\nonumber\\
&&+\frac{1-2r}{9216\pi^2}\langle\frac{\alpha_{s}GG}{\pi}\rangle\int_{x_i}^{1}dx(1-x)^2(2x-17)({s-\tilde{m}_c^2})^2\nonumber\\
&&+\frac{(2-r)m_s\langle\bar{q}q\rangle}{288}\langle\frac{\alpha_{s}GG}{\pi}\rangle\int_{x_i}^{1}dx(4x-5)\nonumber\\
&&+\frac{(1-2r)m_s\langle\bar{s}s\rangle}{384}\langle\frac{\alpha_{s}GG}{\pi}\rangle\int_{x_i}^{1}dx(4x-5)\nonumber\\
&&-\frac{1}{384\pi^2}\langle\frac{\alpha_{s}GG}{\pi}\rangle\int_{x_i}^{1}dx(1-x)^2[-13-8x+r(14+x)]({s-\tilde{m}_c^2})^2\nonumber\\
&&+\frac{m_s\langle\bar{s}s\rangle}{288}\langle\frac{\alpha_{s}GG}{\pi}\rangle\int_{x_i}^{1}dx[-7+20x+r(8-19x)]\,,
\end{eqnarray}

\begin{eqnarray}
\rho^{A,1}_{\frac{5}{2},\Xi_c}(s)&=&\frac{14+2r}{1152\pi^2}\langle\frac{\alpha_{s}GG}{\pi}\rangle\int_{x_i}^{1}dxx(1-x)^4({s-\tilde{m}_c^2})^2\nonumber\\
&&+\frac{(11-7r)m_s\langle\bar{s}s\rangle}{72}\langle\frac{\alpha_{s}GG}{\pi}\rangle\int_{x_i}^{1}dxx(1-x)^2\nonumber\\
&&-\frac{(2-r)m_s\langle\bar{q}q\rangle}{36}\langle\frac{\alpha_{s}GG}{\pi}\rangle\int_{x_i}^{1}dxx(1-x)^2\,,
\end{eqnarray}

\begin{eqnarray}
\rho^{A,0}_{\frac{5}{2},\Xi_c}(s)&=&\frac{1}{192\pi^2}\langle\frac{\alpha_{s}GG}{\pi}\rangle\int_{x_i}^{1}dx(1-x)^4({s-\tilde{m}_c^2})^2\nonumber\\
&&+\frac{(5-4r)m_s\langle\bar{s}s\rangle}{72}\langle\frac{\alpha_{s}GG}{\pi}\rangle\int_{x_i}^{1}dx(1-x)^2\nonumber\\
&&-\frac{(2-r)m_s\langle\bar{q}q\rangle}{18}\langle\frac{\alpha_{s}GG}{\pi}\rangle\int_{x_i}^{1}dx(1-x)^2\, ,
\end{eqnarray}

\begin{eqnarray}
\rho_{j,\Omega_c}^{A,1}(s)&=&2\rho_{j,\Xi_c}^{A,1}(s)\mid_{m_s \to 2m_s, \langle\bar{q}q\rangle \to\langle\bar{s}s\rangle, \langle\bar{q}g_s\sigma Gq\rangle \to\langle\bar{s}g_s\sigma Gs\rangle}\, ,
\end{eqnarray}

\begin{eqnarray}
\rho_{j,\Omega_c}^{A,0}(s)&=&2\rho_{j,\Xi_c}^{A,0}(s)\mid_{m_s \to 2m_s, \langle\bar{q}q\rangle \to\langle\bar{s}s\rangle, \langle\bar{q}g_s\sigma Gq\rangle \to\langle\bar{s}g_s\sigma Gs\rangle}\, .
\end{eqnarray}

\section*{Acknowledgements}
This work is supported by National Natural Science Foundation, Grant Number  12175068.


\begin{thebibliography}{99}

\bibitem{Omegac-Five-LHCb} R. Aaij et al, Phys. Rev. Lett. {\bf 118} (2017)  182001.

\bibitem{Omegac-Four-Belle} J. Yelton et al, Phys. Rev. {\bf D97} (2018)  051102.

\bibitem{Omegac-Four-LHCb} R. Aaij et al, Phys. Rev. {\bf D104} (2021) L091102.

\bibitem{Omegac-Five-LHCb-New} R. Aaij et al, arXiv:2302.04733 [hep-ex].

\bibitem{Chen-Omega-1P} H. X. Chen, Q. Mao, W. Chen, A. Hosaka, X. Liu and S. L. Zhu, Phys. Rev. {\bf D95} (2017)  094008.
\bibitem{Rosner-Omega-1P} M. Karliner and J. L. Rosner,  Phys. Rev. {\bf D95} (2017)  114012.

\bibitem{KLWang-Omega-1P} K. L. Wang, Y. X. Yao, X. H. Zhong and Q. Zhao,  Phys. Rev. {\bf D95} (2017)  116010.

\bibitem{Mathur-Omega-1P} M. Padmanath and N. Mathur, Phys. Rev. Lett. {\bf 119} (2017) 042001.

\bibitem{WangZhu-Omega-1P} W. Wang and R. L. Zhu, Phys. Rev. {\bf D96} (2017)  014024.

\bibitem{WZG-Omega-1P} Z. G. Wang, Eur. Phys. J. {\bf C77} (2017)  325.

\bibitem{Cheng-Omega-1P-2S}  H. Y. Cheng and C. W. Chiang,  Phys. Rev. {\bf D95} (2017) 094018.

\bibitem{ChenB-Omega-1P-2S}  B. Chen and X. Liu, Phys. Rev. {\bf D96} (2017)  094015.

\bibitem{Azizi-Omega-1P-2S} S. S. Agaev, K. Azizi and H. Sundu, Eur. Phys. J. {\bf C77} (2017)  395.

\bibitem{Azizi-Omega-2S} S. S. Agaev, K. Azizi and H. Sundu, EPL {\bf 118} (2017)  61001.

\bibitem{WZG-Omega-2S} Z. G. Wang, X. N. Wei and Z. H. Yan, Eur. Phys. J. {\bf C77} (2017)  832.

\bibitem{Polyako-Omega-penta} H. C. Kim, M. V. Polyakov and M. Praszalowicz, Phys. Rev. {\bf D96} (2017)  014009.

\bibitem{PingJL-Omega-penta} G. Yang and J. Ping, Phys. Rev. {\bf D97} (2018) 034023.

\bibitem{WZG-Omega-penta} Z. G. Wang and J. X. Zhang, Eur. Phys. J. {\bf C78} (2018) 503.

\bibitem{An-Omega-penta} C. S. An and H. Chen, Phys. Rev. {\bf D96} (2017)  034012.

\bibitem{WZG-WHJ-Omega-penta} H. J. Wang, Z. Y. Di and Z. G. Wang, Commun. Theor. Phys. {\bf 73} (2021) 035201.

\bibitem{Oset-Omega-mole} V. R. Debastiani, J. M. Dias, W. H. Liang and E. Oset, Phys. Rev. {\bf D97} (2018)  094035.

\bibitem{YGL-3185-2S-3327-1D}  G. L. Yu,  M. Yan, Z. Y. Li, Z. G. Wang and  J. Lu, arXiv:2302.11758 [hep-ph].

\bibitem{LiuX-3327-1D} S. Q. Luo and X. Liu, arXiv: 2303.04022 [hep-ph].

\bibitem{WZG-Omegab} Z. G. Wang, Int. J. Mod. Phys. {\bf A35} (2020)  2050043.

\bibitem{Wang-2625-2815} Z. G. Wang,  Eur. Phys. J. {\bf C75} (2015)  359.

\bibitem{WZG-D-wave-lambda}  Z. G. Wang, Nucl. Phys. {\bf B926} (2018) 467.

\bibitem{YGL-1D-CPC} G. L. Yu, Z. G. Wang and X. W. Wang, Chin. Phys. {\bf C46} (2022)  093102.

\bibitem{WZG-2S-CPC}  Z. G. Wang and H. J. Wang, Chin. Phys. {\bf C45} (2021)  013109.

\bibitem{One-gluon-1} A. De Rujula, H. Georgi and S. L. Glashow, Phys. Rev.  {\bf D12} (1975) 147.

\bibitem{One-gluon-2} T. DeGrand, R. L. Jaffe, K. Johnson and J. E. Kiskis, Phys.  Rev.  {\bf D12} (1975) 2060.

\bibitem{WangLDiquark}  Z. G. Wang, Commun. Theor. Phys. {\bf 59} (2013) 451.

\bibitem{Korner-PPNP} J. G. Korner, M. Kramer and D. Pirjol, Prog. Part. Nucl. Phys. {\bf 33} (1994) 787.

\bibitem{LHCb2860}   R. Aaij et al, JHEP {\bf 1705} (2017) 030.

\bibitem{Oka96-SB} Y. Chung, H. G. Dosch, M. Kremer and D. Schall,  Nucl. Phys. {\bf B197} (1982) 55.

\bibitem{Oka96} D. Jido, N. Kodama and M. Oka,  Phys. Rev. {\bf D54} (1996) 4532.

\bibitem{WangPc} Z. G. Wang, Eur. Phys. J. {\bf C76} (2016) 70.

\bibitem{WZG-Negative-P} Z. G. Wang, Eur. Phys. J. {\bf A47} (2011) 81.

\bibitem{WZG-Pcs-Old} Z. G. Wang, Eur. Phys. J. {\bf C76} (2016)  142.

\bibitem{WZG-Pcs-NPB} Z. G. Wang, Nucl. Phys. {\bf B913} (2016) 163.

\bibitem{SVZ79} M. A. Shifman, A. I. Vainshtein and V. I. Zakharov, Nucl. Phys. {\bf B147} (1979) 385; Nucl. Phys. {\bf B147} (1979) 448.

\bibitem{PRT85} L. J. Reinders, H. Rubinstein and S. Yazaki, Phys. Rept. {\bf 127} (1985) 1.

\bibitem{HuangShiZhong} Shi-Zhong Huang, "Free particles and fields of high spins" (in chinese), Anhui peoples Publishing House, 2006.

\bibitem{ChenHX-Omega-D-wave} Q. Mao, H. X. Chen, A. Hosaka, X. Liu and S. L. Zhu, Phys. Rev. {\bf D96} (2017)  074021.

\bibitem{ColangeloReview} P. Colangelo and A. Khodjamirian, hep-ph/0010175.

\bibitem{PDG}  R. L. Workman et al, Prog. Theor. Exp. Phys. {\bf 2022} (2022) 083C01.

\bibitem{Narison-mix} S. Narison and R. Tarrach, Phys. Lett. {\bf 125 B} (1983) 217.

\bibitem{WangTetraquark} Z. G. Wang, Eur. Phys. J. {\bf C74} (2014)  2874.

\bibitem{WangIJMPA-mole} Z. G. Wang,  Int. J. Mod. Phys. {\bf A36} (2021)  2150107.

\bibitem{XinQ-EPJA} Q. Xin and Z. G. Wang, Eur. Phys. J. {\bf A58} (2022)  110.

\bibitem{WangEPJC4260} Z. G. Wang, Eur. Phys. J. {\bf C76} (2016) 387.






\end{thebibliography}
\end{document}